\documentclass{aastex631}
\usepackage{graphbox}
\usepackage{float}
\usepackage{subfigure}
\usepackage{rotating}
\usepackage{graphicx}
\usepackage{comment}
\usepackage{silence}

\begin{document}
\submitjournal{ApJ}
\accepted{by ApJ on January 27, 2024}

\title{Dust and Cold Gas Properties of Starburst HyLIRG-Quasars at z $\sim$ 2.5 }
\shorttitle{Starburst HyLIRG-Quasars at z$\sim$2.5}
\shortauthors{Liu et al.}

\author[0000-0003-1386-3676]{Feng-Yuan Liu}
\affiliation{Chinese Academy of Sciences South America Center for Astronomy (CASSACA), National Astronomical Observatories(NAOC),
20A Datun Road, Beijing 100012, China}
\affiliation{School of Astronomy and Space Science, University of Chinese Academy of Sciences, Beijing 101408, China}
\affiliation{Institute for Astronomy, University of Edinburgh, Royal Observatory, Edinburgh EH9 3HJ, UK}

\author[0000-0002-7928-416X]{Y. Sophia Dai}
\affiliation{Chinese Academy of Sciences South America Center for Astronomy (CASSACA), National Astronomical Observatories(NAOC),
20A Datun Road, Beijing 100012, China}
\correspondingauthor{Y. Sophia Dai}
\email{ydai@nao.cas.cn}

\author[0000-0002-4721-3922]{Alain Omont}
\affiliation{Sorbonne Universit$\acute{e}$, UPMC Universit$\acute{e}$ Paris 6 and CNRS, UMR 7095, Institut d'Astrophysique de Paris, France}

\author[0000-0001-9773-7479]{Daizhong Liu}
\affiliation{Max-Planck-Institut f\"ur Extraterrestrische Physik (MPE), Giessenbachstr. 1, D-85748 Garching, Germany}

\author[0000-0002-0679-2219]{Pierre Cox}
\affiliation{Sorbonne Universit$\acute{e}$, UPMC Universit$\acute{e}$ Paris 6 and CNRS, UMR 7095, Institut d'Astrophysique de Paris, France}

\author[0000-0002-7176-4046]{Roberto Neri}
\author[0000-0001-6971-4851]{Melanie Krips}
\affiliation{Institut de Radioastronomie Millimétrique (IRAM), 300 Rue de la Piscine, 38400 Saint-Martin-d'H$\grave{e}$res, France}

\author[0000-0002-8117-9991]{Chentao Yang} 
\affiliation{Department of Space, Earth and Environment, Chalmers University of Technology, Onsala Space Observatory, 439 92 Onsala, Sweden}

\author[0000-0002-7350-6913]{Xue-Bing Wu}
\affiliation{Department of Astronomy, School of Physics, Peking University, Beijing 100871, People's Republic of China}
\affiliation{Kavli Institute for Astronomy and Astrophysics, Peking University, Beijing 100871, People's Republic of China}

\author[0000-0001-6511-8745]{Jia-Sheng Huang}
\affiliation{Chinese Academy of Sciences South America Center for Astronomy (CASSACA), National Astronomical Observatories(NAOC),
20A Datun Road, Beijing 100012, China}
\affiliation{Harvard-Smithsonian Center for Astrophysics, 60 Garden Street, Cambridge, MA, 02215, USA}

\begin{abstract}
Some high-z active galactic nuclei (AGNs) are found to reside in extreme star-forming galaxies,
such as hyper-luminous infrared galaxies (HyLIRGs), 
with AGN-removed $L_{\rm{IR}}$ of $>10^{13}\,L_{\rm{\odot}}$. 
In this paper, we report NOEMA observations of six apparent starburst HyLIRGs 
associated with optical quasars at $z\sim2-3$ in the Stripe 82 field, 
to study their dust and molecular CO properties.
Five out of the six candidates are detected with CO(4-3) or CO(5-4) emission, 
and four in 2mm dust continuum. 
Based on the linewidth-$L'_{\rm{CO(1-0)}}$ diagnostics, 
we find that four galaxies are likely unlensed or weakly lensed sources.
The molecular gas mass is in the range of $\mu M_{\rm{H_2}}\,\sim0.8-9.7\times10^{10}  M_{\odot}$ (with $\alpha = 0.8\, M_{\odot}\,(\rm{K\,km\,s^{-1}\,pc^2})^{-1}$ 
and $\mu$ is the unknown possible gravitational magnification factor).
We fit their SEDs,
after including the observed 2mm fluxes and upper limits,
and estimate their apparent (uncorrected for possible lensing effect)
star formation rates ($\mu$SFRs) to be $\sim400-2500$ $M_{\rm{\odot}}\,\rm{yr^{-1}}$
with depletion time of $\sim20-110$ Myr.
We notice interesting offsets, of $\sim10-40$ kpc spatially 
or $\sim1000-2000$ km$\,$s$^{-1}$ spectroscopically,
between the optical quasar and the mm continuum or CO emissions.
The observed velocity shift is likely related to the blueshifted broad-emission-line region of quasars,
though mergers or recoiling black holes are also possible causes,
which can explain the spatial offset and the high intrinsic SFRs in the HyLIRG-quasar systems. 
\end{abstract}

\keywords{HyLIRGs, CO emission, Millimeter interferometry, Cosmic noon}

\section{Introduction}
\label{sec:intro}

Active galactic nuclei (AGNs) represent a crucial phase in the evolution of supermassive black holes (SMBHs),
and may strongly influence the evolution of their host galaxies \citep{2012ARA&A..50..455F, 2013ARA&A..51..511K}. 
Observations have shown that the powerful phase of AGN, i.e. quasars, influences host galaxies through either radiation pressure \citep[e.g.][]{1993ApJ...402..441L, 1995ApJ...451..510S} or AGN wind \citep[e.g.][]{1991ApJ...373...23W, 2003MNRAS.345..705P, 2012MNRAS.422L...1T} 
during the so-called "quasar mode" feedback.
This is consistent with the evolutionary model of quasars by \citet{1988ApJ...325...74S}, 
where quasars develop from dusty ultra-luminous infrared galaxies (ULIRGs) with infrared (IR) luminosity $L_{\rm{IR}}>10^{12}L_{\odot}$.
As the AGN feedback swepdf the gas and dust in the core region,
the central AGN gets exposed in the line of sight and appears to be a type-I broad-emission-line quasar.
The star formation (SF) in the host galaxy is suppressed during the process.

This model indicates a transitional stage in quasar evolution
where a quasar coexists with a large amount of IR-luminous galactic dust.
In observation, there are 10-30\% quasars with bright submillimeter/far-infrared (SFR) excesses \citep[e.g.][]{2012ApJ...753...33D, 2018MNRAS.478.4238D, 2015ApJ...811...58M, 2016ApJ...824...70D}.
Some of them have the most IR-luminous host galaxies, 
i.e., starburst hyper-luminous infrared galaxies (HyLIRGs) with starburst-dominated $L_{\rm{IR}}>10^{13}L_{\odot}$.
With IR-traced SF rate (SFR) of $\gtrsim 10^3\,M_{\rm{\odot}}\,\rm{yr^{-1}}$ \citep[e.g.][]{2012ApJ...761..139C, 2013ApJ...772..137I, 2013MNRAS.429L..55B}, 
they link the most powerful AGNs and the most extreme SF activities in the host galaxy.

However, the luminosity of these galaxies is questionable because of their potential gravitational magnification.
Many apparent HyLIRGs have been found to be the results of lensing by large-area millimeter/submillimeter surveys \citep[e.g.][]{2010ApJ...719..763V, 2010Sci...330..800N, 2013ApJ...762...59W, 2013ApJ...779...25B,2018A&A...620A..61C} 
and follow-up studies \citep[e.g.][]{2017A&A...608A.144Y,2018MNRAS.481...59Z}.
For instance, the $Herschel$ Astrophysical Terahertz Large Area Survey \citep[H-ATLAS,][]{2010PASP..122..499E} revealed $\sim$1000 strongly lensed sources \citep{2012ApJ...749...65G}.
After correcting for the lensing effect, many sources turn out to be of lower IR luminosities, 
thus no longer HyLIRGs.
As a case, \citet{2016ApJ...829...21T} found that the apparent HyLIRG HATLAS J132427 
is an intrinsic ULIRG, 
after correcting for the magnification factor of five. 
Another apparent HyLIRG SDP.81 has 
a magnification factor of $\sim$18,
reconstructed with ALMA data, 
and is indeed a ULIRG with an SFR of $\sim100 M_{\rm{\odot}}\,\rm{yr^{-1}}$ \citep{2020MNRAS.494.5542R}.

The high apparent IR luminosity can also be an effect of a collection of IR-bright sources.
Luminous infrared galaxies (LIRGs, $L_{\rm{IR}} > 10^{11} L_{\rm{\odot}}$) have been extensively observed in the merging process
\citep[e.g.,][]{2003ApJ...599...92C, 2006ApJ...640..228T, 2010MNRAS.405..219B, 2010ApJ...724..233E, 2011MNRAS.412.1913I, 2013ApJ...772..137I, 2011ApJ...733L..11R},
which is also predicted by simulations
\citep[e.g.,][]{2008MNRAS.391..420S, 2010MNRAS.401.1613N, 2011ApJ...743..159H, 2012MNRAS.424..951H, 2019MNRAS.488.2440M}.
State-of-the-art telescopes with sub-arcsecond resolution powers,
such as the Atacama Large Millimeter/submillimeter Array (ALMA) and the Very Large Array (VLA), 
have resolved some apparent HyLIRGs into multiple sources, 
confirming the resolved sources to be LIRGs or ULIRGs instead.
For example,
\citet{2013Natur.498..338F} resolved two merging ULIRGs in the apparent HyLIRG
1HERMES S250 J022016.5–060143, 
which used to be considered as an unusually bright HyLIRG 
in the $Herschel$ Multi-tiered Extragalactic Survey \citep[HerMES,][]{2012MNRAS.424.1614O}.

CO observations are crucial to study the physical properties of these galaxies.
Firstly, it traces the immediate star-forming material in the host galaxy.
This is connected to the feedback of AGN,
as controversial results have been reported on whether they drive out gas or accelerate star formation efficiency
\citep[SFE, e.g.,][]{2019ApJ...879...41K, 2021A&A...645A..33B}.
Besides, it can reveal the direct feedback from quasars in the form of galactic scale outflows, 
which are exhibited as broad line wings exceeding a velocity of 500 km s$^{-1}$ \citep[e.g.][]{2010A&A...518L.155F, 2012A&A...543A..99C}.
Finally, \citet{2012ApJ...752..152H} has found that strongly lensed galaxies can be distinguished by CO line emission.
Thus, we are able to estimate the lensing property of galaxies under limited resolution.

In this work,
we conduct NOrthern Extended Millimeter Array (NOEMA) observations of the mid-J CO rotational emission (J=4-3 or J=5-4) and 2mm dust continuum in six starburst apparent HyLIRG-quasars at $z\sim2.5$.
The sample is selected from the quasar catalog of the Sloan Digital Sky Survey (SDSS) in the Stripe 82 field, with apparent HyLIRG-level IR luminosity from $Herschel$ observations \citep{2016ApJ...824...70D}.
We use the millimeter observations to probe the molecular gas and dust properties in these sources and to identify if they are intrinsic or lensed starburst HyLIRG-quasars.

The paper is organized as follows: 
In section 2, we describe the sample selection, the observations, and the data reduction process.
In section 3, the observational results are presented, including the continuum
and CO emission morphology and properties, 
and the optical-to-mm spectral energy distributions (SEDs) of our sample.
In section 4, we discuss the sample's location in the HyLIRG diagnostics, 
the spatial and velocity offsets of the sample between various tracers,
and the estimated SFR and depletion time, 
followed by a summary in section 5.

\newpage
\movetabledown=2.2in
\begin{rotatetable}
\begin{deluxetable*}{cccccccccccc}
\tablecaption{Source properties \label{tab:prop}}
\tablecolumns{7}
\tablewidth{0pt}
\tablehead{
\colhead{Source name} &
\colhead{Herschel name\tablenotemark{b}} &
\multicolumn{2}{c}{SDSS coordinates\tablenotemark{a}} &
\multicolumn{2}{c}{Herschel coordinates\tablenotemark{b}} &
\colhead{$S_{\rm{250\mu m}}$\tablenotemark{b}} &
\colhead{$S_{\rm{350\mu m}}$\tablenotemark{b}} &
\colhead{$S_{\rm{500\mu m}}$\tablenotemark{b}} &
\colhead{log$\mu L_{\rm{IR}}$\tablenotemark{c}} & 
\colhead{log$M_{\rm{BH}}$\tablenotemark{d}} & 
\colhead{Redshift\tablenotemark{e}} \\
\colhead{} & \colhead{} & \colhead{$\alpha_{2000}$} & \colhead{$\delta_{2000}$} & 
\colhead{$\alpha_{2000}$} & \colhead{$\delta_{2000}$} & \colhead{(mJy)} & \colhead{(mJy)} & \colhead{(mJy)} & 
\colhead{($L_{\rm{\odot}}$)} 
& \colhead{($M_{\rm{\odot}}$)} 
& \colhead{}
}
\startdata
DW001 & J0111.09-0038.8 & 01:11:05.56 & -00:38:56.35 & 01:11:05.58 & -00:38:54.50 & 40.6$\pm$11.2 & 44.3$\pm$10.9 & 48.0$\pm$11.9 & 13.0$\pm$0.14 & 9.7$\pm$0.09 & 2.8617$\pm$0.0003 \\
DW002 & J0134.04+0039.6 & 01:34:02.83 & 00:39:44.16 & 01:34:02.92 & 00:39:41.60 & 54.2$\pm$11.4 & 83.1$\pm$11.4 & 65.1$\pm$12.4 & 13.1$\pm$0.09 & 9.2$\pm$0.37 & 2.5687$\pm$0.0003 \\
DW003 & J0148.15-0010.2 & 01:48:09.64 & -00:10:17.85 & 01:48:09.46 & -00:10:14.90 & 81.7$\pm$10.9 & 73.5$\pm$11.0 & 76.1$\pm$12.1 & 13.0$\pm$0.06 & 9.8$\pm$0.02 & 2.1528$\pm$0.0004 \\
DW004  & J0156.72+0036.8 & 01:56:43.81 & 00:36:48.70 & 01:56:43.62 & 00:36:47.30 & 35.6$\pm$10.2 & 36.3$\pm$10.3 & 52.1$\pm$11.0 & 13.1$\pm$0.14 & 8.5$\pm$0.09 & 2.0144$\pm$0.0008 \\
DW005 & J0206.76+0105.1 & 02:06:46.34 & 01:05:06.40 & 02:06:46.31 & 01:05:05.57 & 78.8$\pm$11.6 & 82.5$\pm$11.0 & 68.1$\pm$12.3 & 13.0$\pm$0.05 & 9.4$\pm$0.02 & 2.2665$\pm$0.0003 \\
DW006 & J0212.30+0044.9 & 02:12:18.62 & 00:44:56.50 & 02:12:18.50 & 00:44:55.58 & 68.1$\pm$10.7 & 77.5$\pm$10.2 & 68.7$\pm$11.0 & 13.1$\pm$0.06 & 9.7$\pm$0.02 & 2.8664$\pm$0.0009 
\enddata
\tablenotetext{a}{For DW001-DW006, 
if multiple SDSS/BOSS spectra are available, 
the observations with the closest observation date to the NOEMA observations are used (i.e. MJD of 55481, 58107, 58098, 52933, 58079, and 56979, respectively).}
\tablenotetext{b}{The $Herschel$ coordinates and flux densities at 250$\mu$m ($S_{\rm{250\mu m}}$), 350$\mu$m ($S_{\rm{350\mu m}}$), and 500$\mu$m ($S_{\rm{500\mu m}}$) are taken from the HerS catalog \citep[][]{2014ApJS..210...22V}.}
\tablenotetext{c}{Integrated between 8-1000 $\mu$m derived from SED fitting by \citet{2016ApJ...824...70D} considering only the starburst grey-body (i.e. cold dust) component.
Note that we have derived improved values for these luminosities, $\mu L_{\rm{IR,SB}}$, in Table \ref{tab:sed}.}
\tablenotetext{d}{The virial BH masses estimated from broad C IV lines \citep{2011ApJS..194...45S, 2016ApJ...824...70D}.}
\tablenotetext{e}{SDSS spectroscopic redshifts, which are dominated by broad C IV lines \citep[see][and discussion in Section \ref{sec:disoff}]{2011ApJS..194...45S, 2020ApJS..250....8L}.}
\end{deluxetable*}
\end{rotatetable}

Throughout this paper, we adopt a $\Lambda$CDM cosmology with $\rm{\Omega_m} = 0.32$, $\rm{\Omega_{\Lambda}} = 0.68$, $\rm{\Omega_{k} = 0}$, and $\rm{H_0 = 67\, km\ s^{-1}\ Mpc^{-1}}$ \citep[][]{2020A&A...641A...6P}. 
In addition, we adopt the Chabrier initial mass function \citep[IMF,][]{2003PASP..115..763C}
for our SFR estimates.
\footnote {$\mu SFR\,\, (M_{\odot}\,\rm{yr^{-1}}) = 1.2\times10^{-10}\, \mu L_{\rm{IR, SB}}\,\, (L_{\rm{\odot}})$,
where $\mu L_{\rm IR,SB}$ is the AGN-removed, pure starburst IR luminosity \citep[converted from][]{1998ApJ...498..541K}, after applying a correction factor of 0.7 for the Chabrier IMF \citep[]{2008MNRAS.385..147D}.}

\section{Sample selection and observations}
\subsection{Sample selection}

\begin{figure}[H]
\centering  
\includegraphics[width=0.8\textwidth]{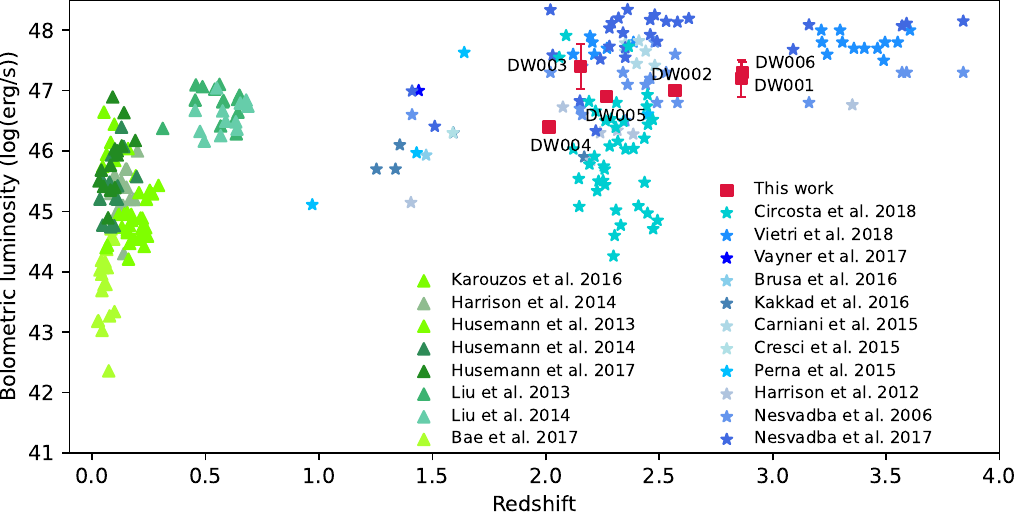}
\caption{Redshifts of selected sources versus their bolometric luminosities.
Red filled squares are the six HyLIRG-quasars with bolometric luminosities from \citet{2016ApJ...824...70D}. The triangles represent AGN systems at z $<$ 1 \citep{2017ApJ...837...91B, 2016ApJ...819..148K, 2014MNRAS.441.3306H, 2013A&A...549A..43H, 2014MNRAS.443..755H, 2017MNRAS.470.1570H, 2013MNRAS.430.2327L, 2014MNRAS.442.1303L}.
Filled stars represent high-redshift AGN \citep{2018A&A...617A..81V, 2017ApJ...851..126V, 2016A&A...588A..58B, 2016A&A...592A.148K, 2015A&A...580A.102C, 2015ApJ...799...82C, 2015A&A...583A..72P, 2012MNRAS.426.1073H, 2010MNRAS.402.2211A, 2006ApJ...650..693N, 2017A&A...600A.121N}.
The six sources we selected have comparable bolometric luminosities 
compared to known sources at similar redshifts.}
\label{fig:boloz}
\end{figure}

We select six apparent starburst HyLIRG-quasars from the catalog in \citet{2016ApJ...824...70D}.
This sample was selected in the Stripe 82 field from the SDSS quasar catalogs \citep{2010AJ....139.2360S, 2011ApJS..194...45S, 2014A&A...563A..54P}
, which were pre-selected to be brighter than $M_i = -22.0$ 
and have at least one optical line with full width at half-maximum (FWHM) larger than 1000 km s$^{-1}$ (Type 1).
These quasars were then cross-matched with the $Herschel$ Stripe 82 Survey \citep[HerS,][]{2014ApJS..210...22V} by \citet{2016ApJ...824...70D}, 
and 207 showed $Herschel$ SPIRE detections
at 250, 350, and 500 $\rm{\mu m}$.
These wavelengths cover the spectral regions close to the peak 
of the cold dust emission at $z\sim2.5$ and 
thus can better constrain the cold dust properties in the SED fitting.
AGN-subtracted IR luminosity $L_{\rm{IR, 8-1000\mu m}}$ 
was then calculated based on a grey-body dust component \citep{2016ApJ...824...70D}.

We then selected sources with $L_{\rm{IR, 8-1000\mu m}} > 10^{13} L_{\rm \odot}$ (Table \ref{tab:prop}), 
i.e., apparent starburst HyLIRGs.
To avoid possible gravitational lensing and blending issues,
we further required the targets to be point sources in the SDSS images, 
without any close companion within 5\,\arcsec, which is slightly larger than 
the NOEMA resolution ($\sim4\,\arcsec$) (D configuration).
This way, we selected six starburst HyLIRG-quasars from the \citet{2016ApJ...824...70D} catalog,
namely DW001 to DW006.
All of the selected sources have spectroscopic redshifts between 2 and 3, 
corresponding to the peaks of cosmic evolution for both star formation and AGN accretion \citep{2020ARA&A..58..661F}.
Some of them have multi-epoch observations by SDSS,
including those from the Baryon Oscillation Spectroscopic Survey \citep[BOSS,][]{2011AJ....142...72E} and
the Extended Baryon Oscillation Spectroscopic Survey \citep[eBOSS,][]{2016AJ....151...44D}.
For the convenience of later comparison with millimeter observations, 
we use optical properties derived from the spectra taken at the closest time to our NOEMA observations.

The physical properties of the six quasars are listed in Table \ref{tab:prop}.
The virial black-hole mass listed in Table \ref{tab:prop} was based on the broad CIV lines
 \citep{2011ApJS..194...45S},
with a typical of $M_{\rm{BH}} \gtrsim 10^9 M_{\odot}$ (except DW004), 
placing them among the most massive quasars.
The bolometric luminosities ($L_{\rm bol}$) of the selected sources are $10^{46.4-47.4}$ erg s$^{-1}$.
Figure \ref{fig:boloz} shows their positions in the redshift-$L_{\rm bol}$ plane. 
Our sources have comparable bolometric luminosities with AGNs at similar redshifts from the literature.

\subsection{Observations and data reduction}
 \label{sec:obs}
We observed our targets with NOEMA (S20BT, PI: Dai) in the 2 mm band 
with 10 antennas on June 6, 15, 19, and September 17, 2020 (Table\,\ref{tab:obs}).
The compact D configuration was chosen to achieve the highest sensitivity.

\begin{figure}[H]
\centering  

\subfigure{}{
\label{dw005a}
\includegraphics[align=c, width=0.27\textwidth]{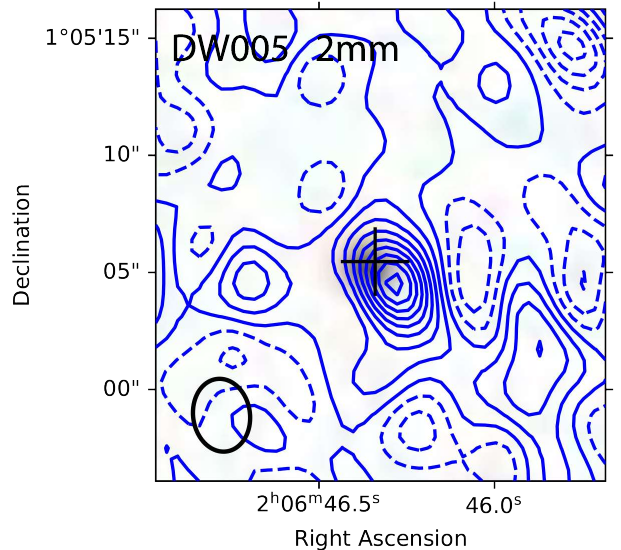}
\includegraphics[align=c, width=0.29\textwidth]{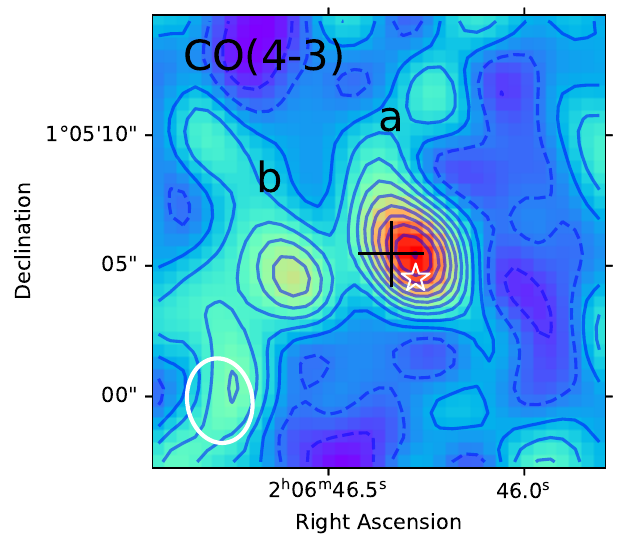}
\includegraphics[align=c, width=0.41\textwidth]{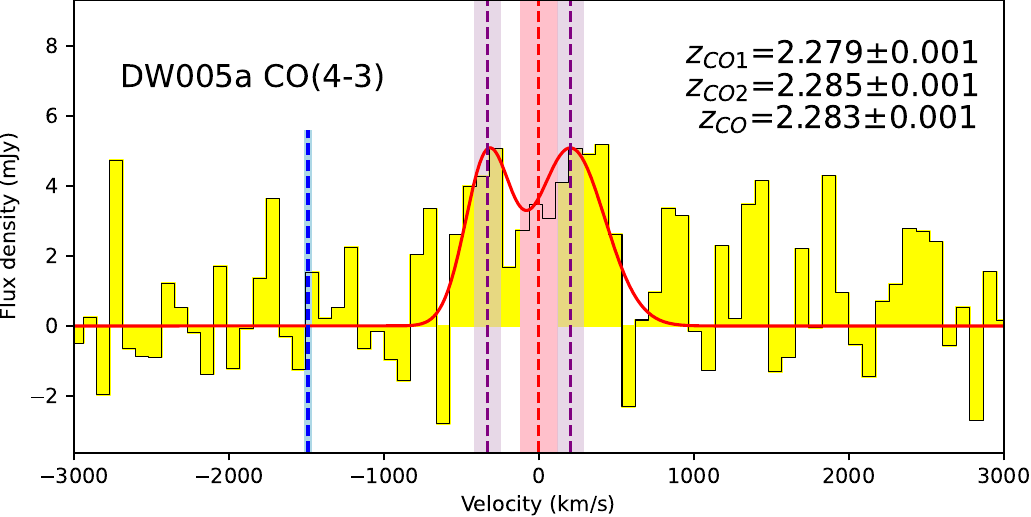}
}

\subfigure{}{
\label{dw005b}
\includegraphics[align=c, width=0.27\textwidth]{blank.pdf}
\includegraphics[align=c, width=0.29\textwidth]{blank.pdf}
\includegraphics[align=c, width=0.41\textwidth]{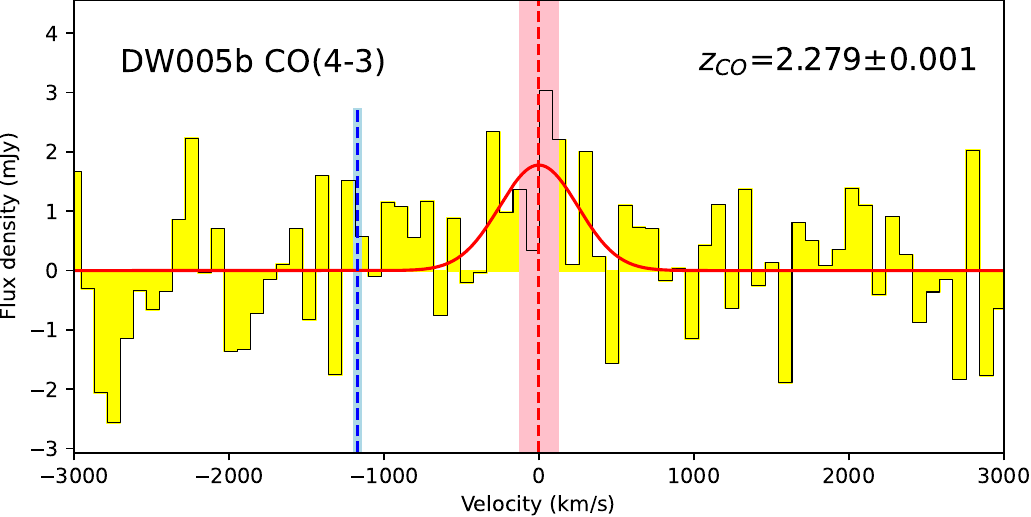}
}

\subfigure{}{
\label{dw006}
\includegraphics[align=c, width=0.27\textwidth]{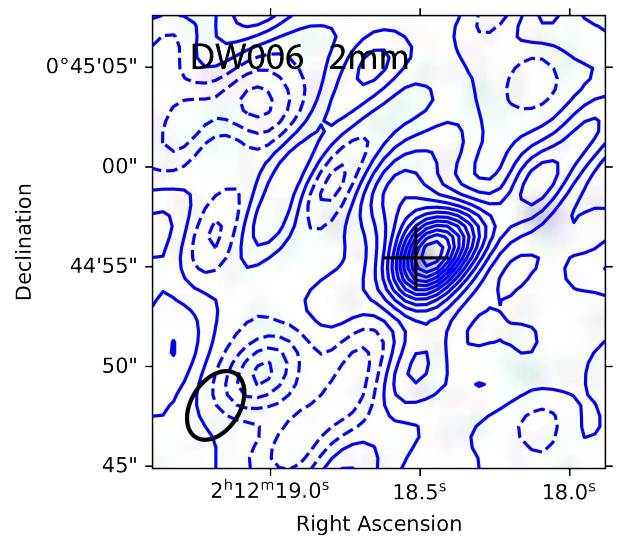}
\includegraphics[align=c, width=0.29\textwidth]{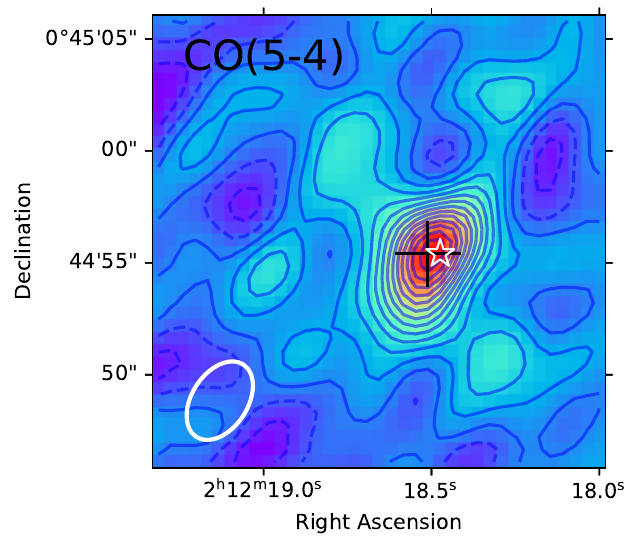}
\includegraphics[align=c, width=0.41\textwidth]{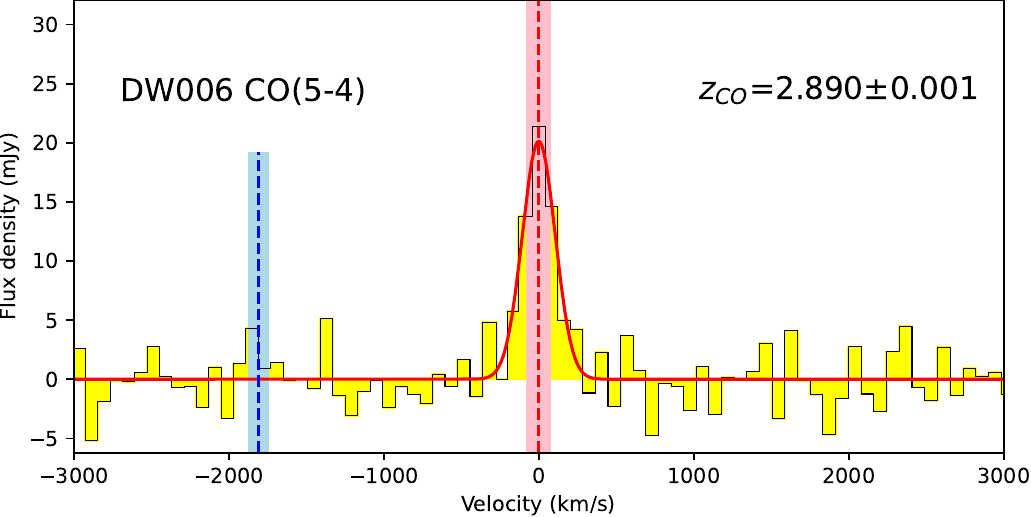}
}

\caption{The 2mm continuum (left) and CO (middle) maps, and CO spectra (right) of 
DW001 to DW006 (from the top).
In both maps, the solid contours represent positive values from 0, 
while the dashed contours are negative values from -1$\sigma$,
with a 1$\sigma$ spacing (see Table \ref{tab:fit} for the $\sigma$ values).
The beam sizes are plot in the bottom left corners.
The quasar's SDSS positions are marked with black crosses,
and the white stars in the CO maps 
mark the peak positions of the continuum. 
Note that for DW003, the peak of the dust continuum (DW003b) is significantly 
offset from the quasar's position (DW003a),
while the CO emission comes mostly from the optical quasar (DW003a), 
though with a 4.3$\arcsec$ offset.
The right column shows the continuum-subtracted CO spectra, 
with a velocity resolution of $\sim$80 km\,s$^{-1}$.
Red curves are the Gaussian fits for the lines. 
The vertical red dotted lines mark the fitted CO line centers,
with their 1$\sigma$ errors in pink shadows (Table \ref{tab:fit}). 
The vertical blue dotted lines and shades mark the expected CO frequency and 1$\sigma$ dispersion
based on the optical quasars' redshifts. 
For DW005a, the purple lines mark the positions for the two peaks, 
and the red vertical line represent the flux-weighted central position (see Table \ref{tab:fit}).}
\label{fig:obs}
\end{figure}

We used the $Herschel$ coordinates of the sources as the phase centers (Figure \ref{fig:obs}).
The targets were observed with the PolyFix correlator with two sidebands of 7.744GHz bandwidths.  
At $z=$ 2 -- 3, the equivalent velocity coverage is $\sim$14000-16000 km\,$s^{-1}$ in each sideband.
We adjust the  $\sim$140-160 GHz spectral windows (Table \ref{tab:obs}) to cover the $^{12}$CO(4-3) (rest frequency 461.040 GHz) for $z<2.5$ targets, and $^{12}$CO(5-4) (rest frequency 576.267 GHz) for $z>2.5$ targets.
The expected CO lines are set close to the center of one 3.8 GHz baseband.
The native channel width was 2 MHz, 
and resampled to $\sim$20 MHz during the data calibration process, 
corresponding to $\sim$40 km/s.

\begin{deluxetable*}{ccccccc}[ht]
\tablecaption{Observations \label{tab:obs}}
\tablecolumns{7}
\tablewidth{0pt}
\tablehead{
\colhead{Tuning set} &
\colhead{Target name} & 
\colhead{Observation Date} & 
\colhead{Exposure time per source} & 
\colhead{Rest frequency\tablenotemark{a}}
&\colhead{Baseline}
& \colhead{PWV\tablenotemark{b}} \\
\colhead{} & \colhead{} & \colhead{} &
\colhead{(Hour)} 
& \colhead{(GHz)} 
& \colhead{(m)}
& \colhead{(mm)}
}
\startdata
S20BT001 & DW001, DW002, DW003, DW006 & 2020\,Jun\,06 & 0.3 & 148.256 & 24-176 & 6-8 \\
 &  & 2020\,Jun\,19 & 0.5 &  & & 2-3\\
S20BT002 & DW004, DW005 & 2020\,Jun\,17 & 0.8 & 141.500 & 24-176 &5-9\\
\enddata
\tablenotetext{a}{Defined as the zero velocity in the lower sideband. For DW001, DW002, DW003, and DW006, the frequency coverage is 135.692GHz to 143.436GHz (the lower sideband) and 151.180GHz to 158.924GHz (the upper sideband). For DW004 and DW005, the frequency coverage is 144.576GHz to 152.320GHz (the lower sideband) and 160.064GHz to 167.808GHz (the upper sideband).}
\tablenotetext{b}{Precipitable water vapor.}
\end{deluxetable*}

We took advantage of the track sharing mode, 
and grouped our sources into two frequency tuning set
(DW001, DW002, DW003, and DW006 in S20BT001
and DW004 and DW005 in S20BT002).
The observations on June 15 of S20BT001 was not used
due to bad data quality. 
For S20BT001 and S20BT002, 
the phase and amplitude calibrators were J0122-003 and 0215+015, respectively;
the radio frequency (RF) calibrators were 3C454.3 and 3C84, respectively;
and the flux calibrators were 1749+096, MWC349, 2010+723, 0106+013, 0215+015.

We used the GILDAS CLIC and MAPPING packages 
to manually calibrate and reduce the data. 
The dust continuum was extracted from the uv tables and 
the flux was derived with 2-D Gaussian fit in the uv plane.
A $5\sigma$ upper limit was assigned for non-detections,
where $\sigma$ is the $1\sigma$ dispersion of 
the brightness distribution within a $30''\times30''$ field centered at the source.
The CO fluxes were extracted with 2-D Gaussian fits on the uv plane for each channel, 
generated from the continuum-subtracted uv tables. 
The CO spectra were then constructed by combining the extracted fluxes of every channel.
DW003 and DW005 show two CO-emitting components exceeding $4\sigma$ on the velocity-collapsed images, 
so we used two Gaussian components to extract the fluxes individually onn the uv plane.
On the extracted spectra, we fit Gaussian profiles ($\pm$1000\,km\,s$^{-1}$ around the channel of the peak flux)
and derived the line properties such as the FWHM, 
the peak flux density, and the integrated flux.

To test if the sources are resolved, we also extracted the continuum fluxes with a point source model for comparison.
For DW001 and DW005, the point-source extracted fluxes are consistent with the 2D-Gaussian extracted fluxes (within $1\sigma$) and the FWHMs of the Gaussian profiles are smaller than the beam size. 
Therefore, we treat DW001 and DW005 as unresolved and use the point-source extracted fluxes.
For DW002, DW003b and DW006, the 2D-Gaussian model gives a larger flux and cleaner residual.
The typical fitted FWHM is $\sim4''$,
which is marginally resolved with our beam sizes of $\sim3''-4''$,
thus fluxes measured with the Gaussian profile were used.

Figure \ref{fig:obs} displays the dust continuum emission together with the CO emission images 
that were obtained by collapsing the datacubes within the fitted velocity ranges of the CO emission lines.
We adopted natural weighting for the mapping process, 
set the cleaning threshold to be 50\% of the $1\sigma$ noise of the $30''\times30''$ dirty map, 
and tried various degrees of tapering to all sources.
For the CO emission of DW003, tapering creates larger synthesized beams ($6.2''\times3.7''$), thus concentrating the smeared-out flux and providing a higher peak-flux signal-to-noise ratio (SNR), 
which just exceeds our detection criterion of $5\sigma$ (from $SNR\sim4$ to $\sim6$).
For other sources where the detection is not affected by tapering, 
we kept the original resolution (typical beam size is $3.5\arcsec\times2.5\arcsec$, 
see Table \ref{tab:off}).

\begin{deluxetable*}{lcccccccccc}[ht]
\tablecaption{Observed CO and dust properties \label{tab:fit}}
\tablecolumns{7}
\tablewidth{0pt}
\tablehead{
\colhead{Source} &
\colhead{Targeted} &
\colhead{Frequency} & 
\colhead{Line} & 
\colhead{Line flux} & 
\colhead{Peak flux}&
\colhead{Peak flux}&
\colhead{CO redshift} &
\colhead{Continuum}&
\\
\colhead{name} &
\colhead{line} &
\colhead{center \tablenotemark{a}} & 
\colhead{width} & 
\colhead{} & 
\colhead{density\tablenotemark{b}}&
\colhead{} & 
\colhead{} &
\colhead{flux density\tablenotemark{d}} \\
\colhead{} & \colhead{} &\colhead{(GHz)} & \colhead{(km~s$^{-1}$)} &
\colhead{(Jy~km~s$^{-1}$)} 
& \colhead{(mJy)} 
& \colhead{(Jy/beam\,km\,s$^{-1}$)}
&\colhead{} 
& \colhead{(mJy)}
\\
}
\startdata
DW001 &CO(5-4)& 149.151 & 480$\pm$160 & 2.3$\pm$0.6 & 4.6$\pm$1.1 & 1.5$\pm$0.2 & 2.864$\pm$0.001 & 0.28$\pm$0.05 \\
DW002 &CO(5-4)& 161.422 & 680$\pm$170 & 6.2$\pm$1.4 & 8.7$\pm$1.9 & 1.8$\pm$0.3 & 2.570$\pm$0.001 & 0.96$\pm$0.11  \\
DW003a &CO(4-3)& 145.441 & 430$\pm$110 & 3.1$\pm$0.7 & 6.8$\pm$1.5 & 1.7$\pm$0.3 &2.170$\pm$0.001 & $<0.24$ \\
DW003b & & 145.522 & 270$\pm$100 & 0.6$\pm$0.2 & 2.1$\pm$0.7 & 1.1$\pm$0.3 & 2.168$\pm$0.001& 0.54$\pm$0.11  \\
DW004 &CO(4-3)& - & - & $<1.3$\tablenotemark{c} &- &-& - & $<0.29$\tablenotemark{c} \\
DW005a &CO(4-3)& 140.441 & 530$\pm$120 & 4.6$\pm$0.8 & 5.1$\pm$0.9 & 2.3$\pm$0.3 & 2.283$\pm$0.001 & 0.48$\pm$0.06 \\
DW005b & & 140.592 & 600$\pm$230 & 1.2$\pm$0.4 & 1.8$\pm$0.6 & 1.2$\pm$0.3 &2.279$\pm$0.001 &$<0.32$\tablenotemark{c}\\
DW006 &CO(5-4)& 148.146 & 250$\pm$30 & 5.4$\pm$0.6 & 20.3$\pm$1.9 & 2.9$\pm$0.2 &2.890$\pm$0.001 & 1.27$\pm$0.12 \\
\enddata

\tablenotetext{a}{With a typical error of $\sim$0.04-0.06 GHz. For those that have fitted uncertainty of CO line frequency lower than 0.04 GHz, we use the spectral resolution of 0.04 GHz (corresponding to a velocity resolution of $\sim80$ km\,s$^{-1}$) as their errors.}
\tablenotetext{b}{Corresponding to a spectral resolution of $\sim$80 km\,s$^{-1}$.}
\tablenotetext{c}{Upper limits are derived by 5$\sigma$ background noise times one beam size.}
\tablenotetext{d}{See text for the way the continuum flux density is derived.}
\end{deluxetable*}

\begin{figure}[H]
\centering  

\subfigure{}{
\label{dw001}
\includegraphics[align=c, width=0.27\textwidth]{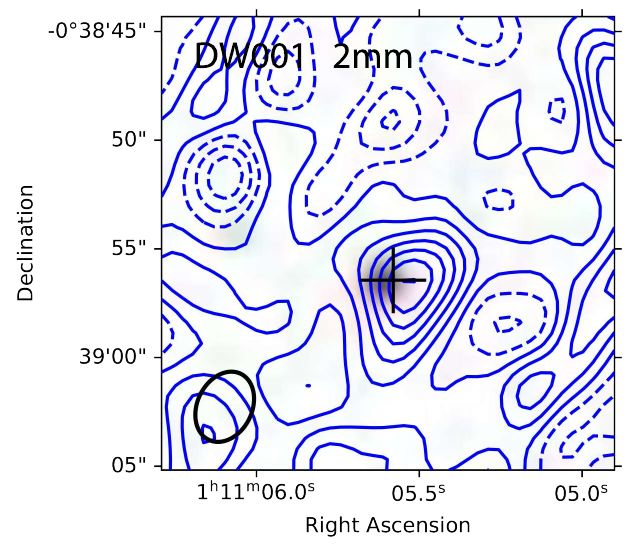}
\includegraphics[align=c, width=0.29\textwidth]{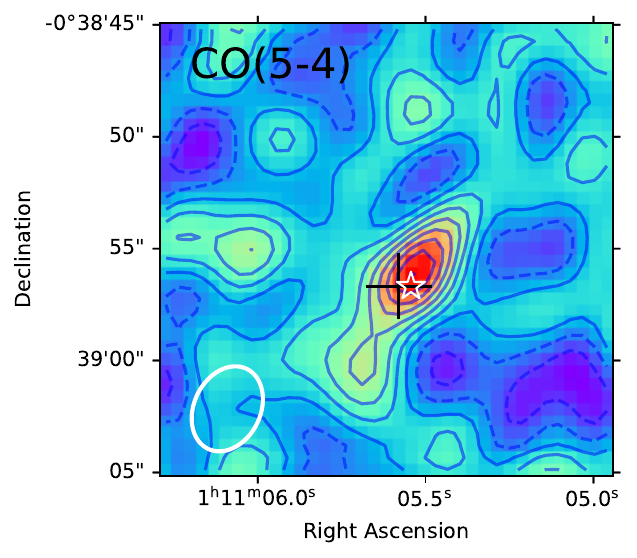}
\includegraphics[align=c, width=0.41\textwidth]{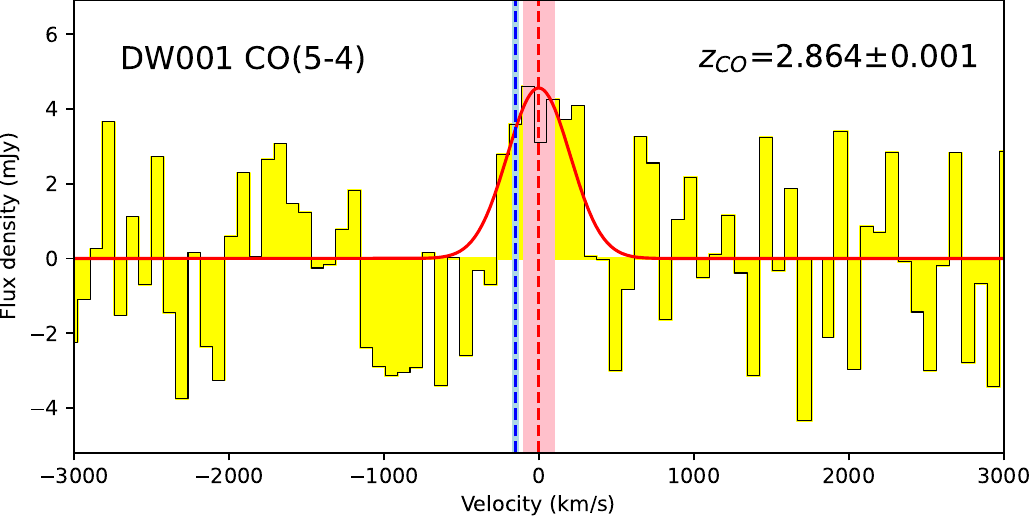}
}

\subfigure{}{ 
\label{dw002}
\includegraphics[align=c, width=0.27\textwidth]{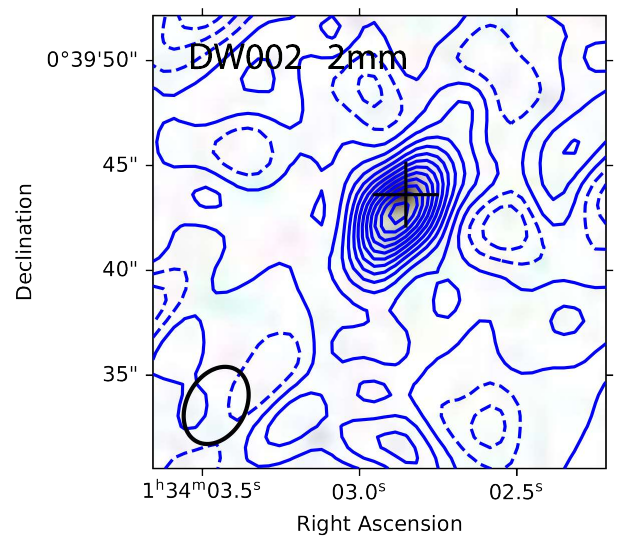}
\includegraphics[align=c, width=0.29\textwidth]{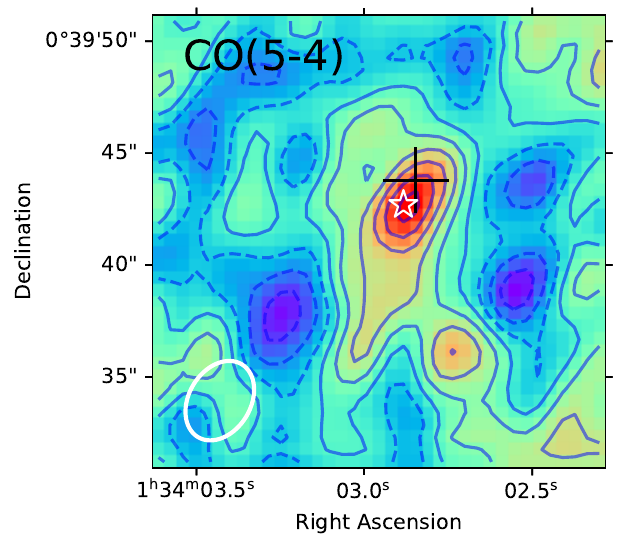}
\includegraphics[align=c, width=0.41\textwidth]{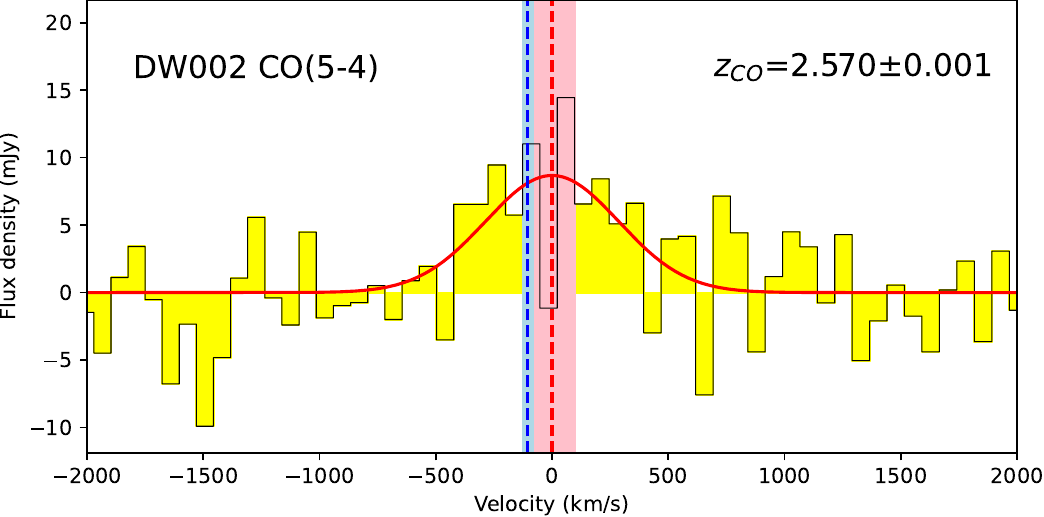}
}

\subfigure{}{
\label{dw003a}
\includegraphics[align=c, width=0.27\textwidth]{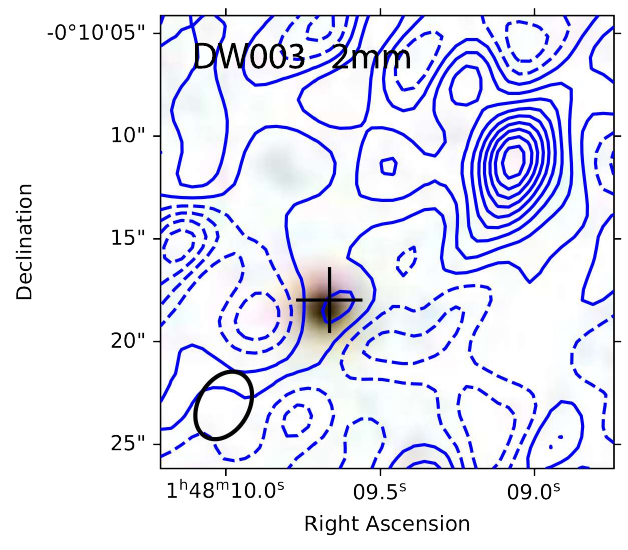}
\includegraphics[align=c, width=0.29\textwidth]{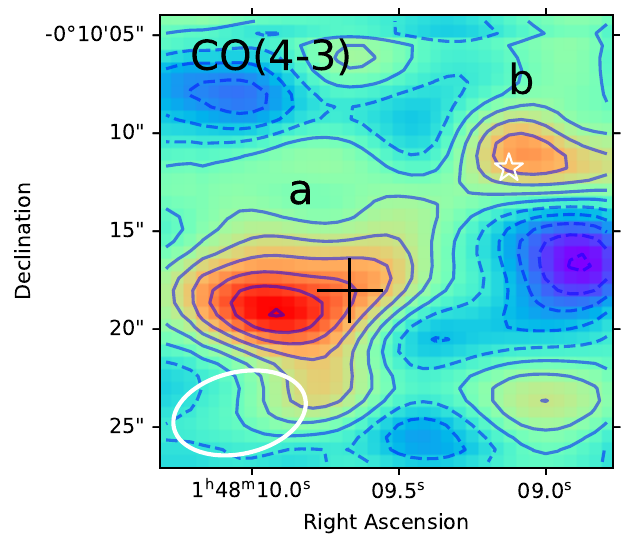}
\includegraphics[align=c, width=0.41\textwidth]{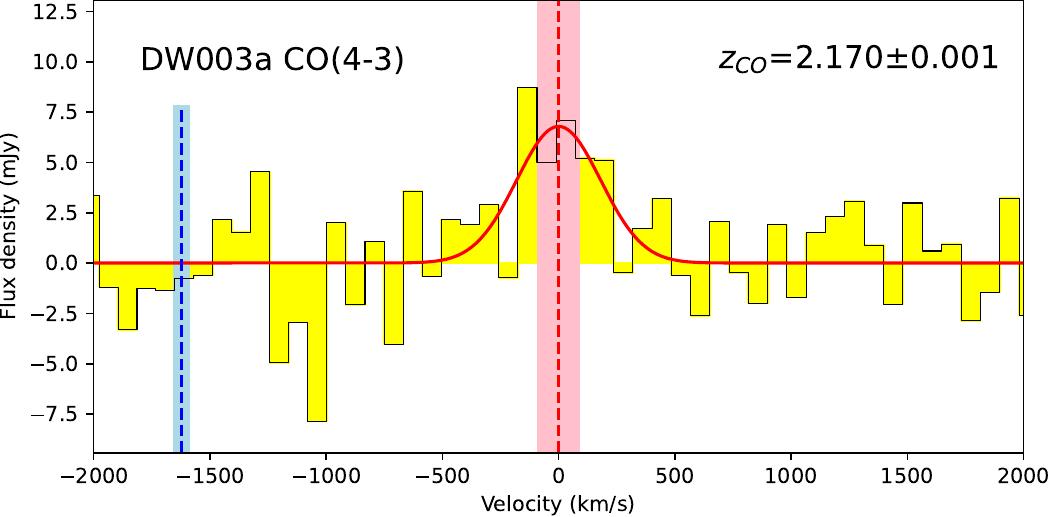}
}
\subfigure{}{
\label{dw003b}
\includegraphics[align=c, width=0.27\textwidth]{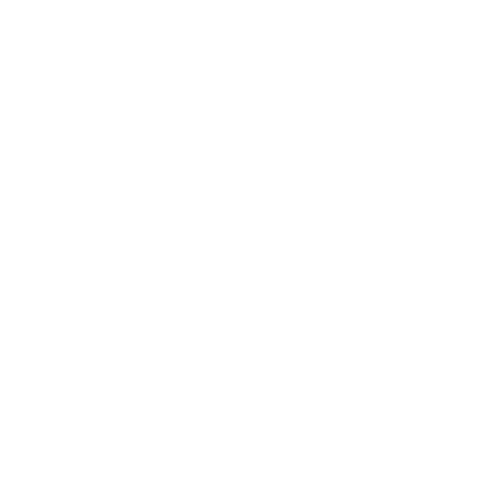}
\includegraphics[align=c, width=0.29\textwidth]{blank.pdf}
\includegraphics[align=c, width=0.41\textwidth]{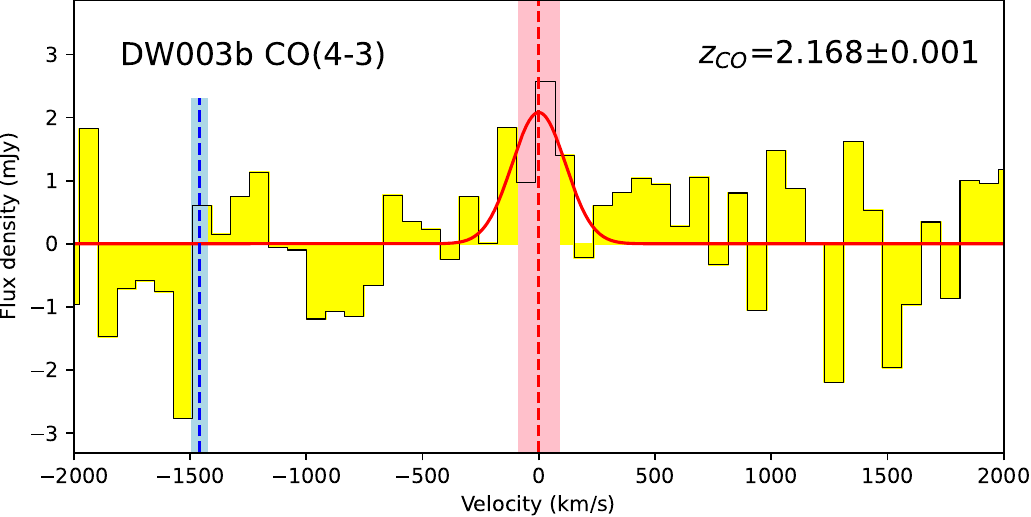}
}

\subfigure{}{
\label{dw004}
\includegraphics[align=c, width=0.27\textwidth]{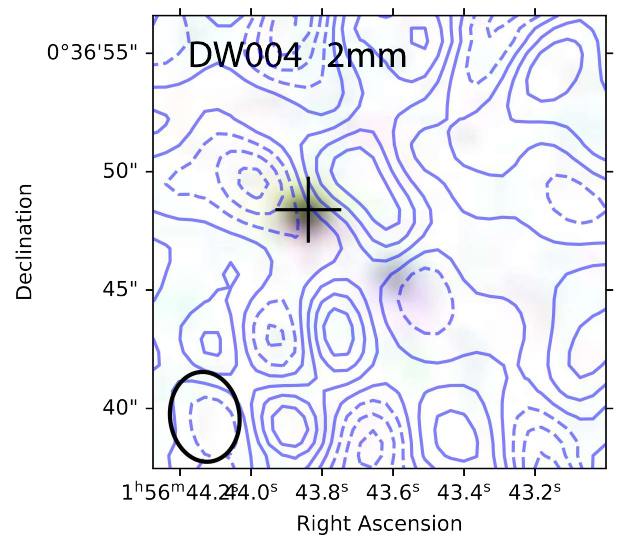}
\includegraphics[align=c, width=0.29\textwidth]{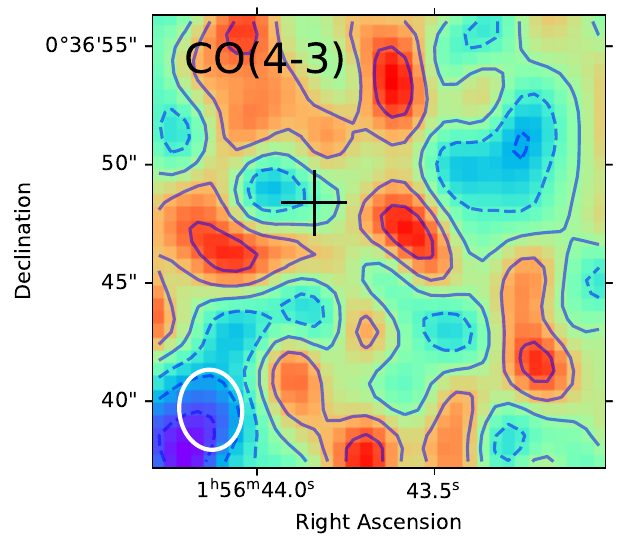}
\includegraphics[align=c, width=0.41\textwidth]{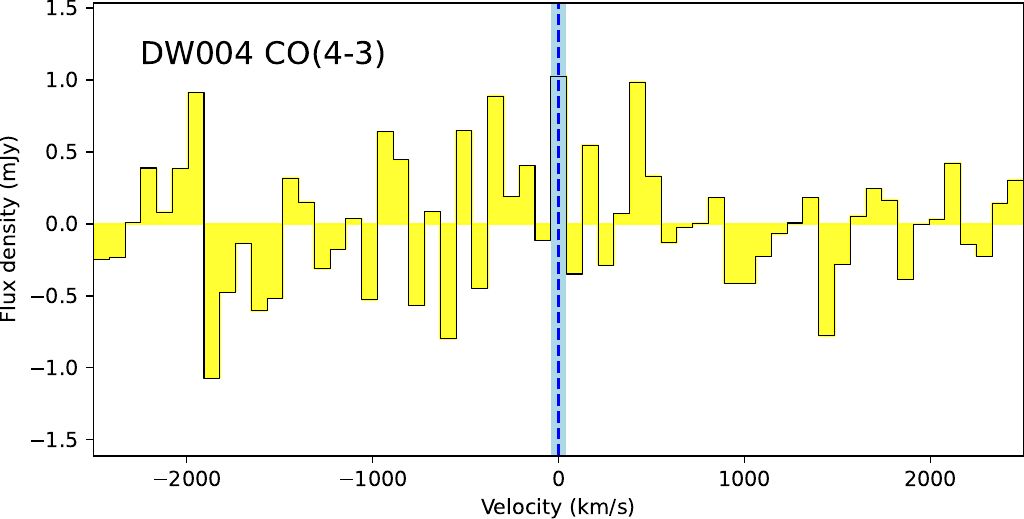} 
}

\end{figure}

\section{Dust Continuum, Molecular Gas Property and Morphology}

\subsection{Dust Continuum} \label{sec:continuum}
We detect the dust continuum (peak flux density $> 5\sigma$)
in four of the six sources: 
DW001, DW002, DW005, and DW006.  
We also detect a $> 5\sigma$ dust continuum $\sim$90 kpc away from 
the DW003 optical position.
The left column of Figure \ref{fig:obs} shows the
continuum contours overlaid on the optical images from SDSS.
The measured dust continuum flux densities are listed in Table \ref{tab:fit}.

We note that the dust continuum 
of DW003 at the optical position (i.e, DW003a) is only of 3.5$\sigma$. 
However, $\sim 10\arcsec$ north-east, 
there is a $>6\sigma$ continuum component (DW003b) without an optical counterpart,
suggesting an optically-obscured submillimeter galaxy (SMG).
Given the tentative CO detection at the same redshift,
DW003b is likely associated with the targeted quasar (Section \ref{sec:spec}).
We note that the 250$\mu$m HerS position is between DW003a and DW003b, 
indicating blending in the FIR dust emission.
For DW004, we detect nothing at 3$\sigma$-level.
For DW005, although the continuum is detected at 7$\sigma$, 
$\sim4.7\arcsec$ east of the quasar,
we find a companion with significant CO emission only (DW005b, 
see Section \ref{sec:spec} and Figure \ref{fig:obs}). 
To calculate the continuum flux upper limits of DW004 and DW005b,
we use the 5$\sigma$ upper limit (see Section \ref{sec:obs}) times the synthesized beam size.
The slightly extended 3$\sigma$ continuum contour of DW006 could be a possible lensing effect, which will be discussed in Section \ref{sec:hylirg}.
We also find spatial offsets and velocity shifts between the optical and mm observations,
which will be described and discussed in Section \ref{sec:offset} and Section \ref{sec:disoff}.

\subsection{Molecular CO Emission}\label{sec:spec}

The central column of Figure \ref{fig:obs} shows the molecular CO emissions,
and the right column is the continuum-subtracted, integrated spectra.
The CO(4-3) or CO(5-4) emission are detected with a peak SNR level of $\sim6-13$
for five of the six sources, except DW004.  
The measured line properties are listed in Table \ref{tab:fit}.

For DW001 and DW002, the CO(5-4) emissions both show a slightly extended structure 
along the major axis of the clean beam. 
Both have a CO emission that aligns well with the 2mm dust continuum. 

For DW003, there was no $>$5$\sigma$ detection within the 
original beam resolution ($4.02\arcsec$$\times$$2.78\arcsec$) so tapering was adopted.
After tapering, a larger beam size (6.16\arcsec$\times$$3.73\arcsec$)
yields a 6$\sigma$ CO(4-3) detection (DW003a), 
which extends to the south-east of the optical quasar position (cross in Figure~\ref{fig:obs}).
This corresponds to a molecular gas reservoir up to $10\arcsec$ (SNR$>$3 region of $\sim$80 kpc).
To the north-west of DW003a, at the location of DW003b,
we also detect a weak CO(4-3) emission of $\sim3\sigma$.
The ratio between continuum flux density and CO peak flux density of DW003b ($\sim0.3$) is higher than other sources ($<0.1$, Table \ref{tab:fit}), which may indicate the existence of an obscured AGN.
DW003a and DW003b have almost identical emission line frequencies,
suggesting that they are in a pair system.
The projected distance is $\sim110$ kpc ($\sim14\arcsec$). 
Similar separations have been reported before \citep[e.g., between NGC7679 and NGC7682 by][]{2021MNRAS.506.5935R},
suggesting possible interaction between the two systems, 
which might have result in the extended CO morphology of DW003a.

DW004 is not detected in CO,
despite integrating through a range of 2000\,km\,s$^{-1}$ around the expected frequency.
The spectrum of DW004 is subtracted using a polygonal aperture covering the phase center, 
which is of similar sizes to the other detected sources ($\sim$10 beams).
We estimate the upper limit of CO(4-3) emission flux
as 5$\sigma$ noise times one beam size.

DW005 have two components.  
The higher SNR CO(4-3) feature aligns well with the peak of the 2mm dust continuum (DW005a). 
The related CO emission shows a double-peak line profile, 
with separation of $\sim650\,\rm{km\,s^{-1}}$.
The double-peaked profile may indicate either disk-like rotation in the system, 
or two distinct velocity components.
In the following analysis, we fit the two lines separately, 
and also calculate an average velocity for DW005a, 
weighted by the fitted fluxes of the two lines.
The average velocity corresponds to a redshift is 2.283$\pm$0.001.
We also find a potential second 
CO emitter (DW005b) $\sim5\arcsec$ east of DW005a (Figure \ref{fig:obs}).
DW005b has an integrated CO flux of $> 3\sigma$,
at an almost identical redshift of the bluer peak of DW005a, 
though not associated with any $>3\sigma$ dust continuum(Section \ref{sec:continuum}).
The different line profiles of the two components suggest that they are not 
likely caused by gravitational lensing.

In DW006, the CO(5-4) line is detected at a $\sim10\sigma$ level
with a relatively narrow width (FWHM $=$ 250 $\rm{km\,s^{-1}}$).
Coincident dust continuum and optical positions,
the strong emission and narrow FWHM are indicative of 
a lensed system as will be discussed in Section~\ref{sec:hylirg}.

\subsection{Positional and Velocity Offsets} \label{sec:offset}

We find common (3 out of 5 detected) positional offsets and velocity shifts between 
the NOEMA continuum emission and SDSS optical observations in our sample, 
as listed in Table \ref{tab:off}. 
However, the moment maps, 
generated using the GILDAS MAPPING package,
do not show any significant signs of velocity components.

For DW001, DW002, DW005 and DW006, 
the peak positions of the 2mm continuum are 
offset from the quasar optical positions by $\sim0.4\arcsec-1.6\arcsec$.
We calculate the spatial uncertainties using
\begin{equation}\label{equ1}
	\delta \theta = \left({\frac{4}{\pi}}\right)^{1/4}\frac{\theta_{\rm{Beam}}}{\sqrt{8\, \rm{ln}2} \times{SNR} _{\rm{peak}}}
\end{equation}
where $\delta \theta$ is the positional uncertainty, 
$\theta_{\rm{Beam}}$ is the cleaned beam size ($3\arcsec-4\arcsec$),
and $SNR_{\rm{peak}}$ is the signal-to-noise ratio of the peak detection on the map \citep{1988ApJ...330..809R}.
All our sources have consistent $Gaia$ \citep{2016A&A...595A...2G, 2021A&A...649A...1G} positions
with the SDSS coordinates, except for DW003 and DW005, whose Gaia positions are 40\,mas north. 
Since 40 mas is negligible compared to the 2mm uncertainties, 
in the following analysis and Table \ref{tab:off}, 
all offsets are calculated based on the SDSS optical positions.
The positional uncertainty $\delta \theta$, 
the NOEMA pointing accuracy of $0.2\arcsec$, 
the NOEMA pixel size of $0.5''$,
and the $SDSS$ positional uncertainty of $<0.1''$ 
together yield a spatial uncertainty of $1.0''$ for DW003b, and $0.5''$ for the other sources.
Since DW003a and DW005b have no significant dust continuum detection, 
in Table~\ref{tab:off} (column 5 and 6) we present their CO flux peak position offsets 
relative to the optical positions.

\begin{deluxetable*}{lcccccccc}[t]
\tablecaption{Source offsets \label{tab:off}}
\tablecolumns{8}
\tablewidth{0pt}
\tablehead{
\colhead{Source name} &
\multicolumn{2}{c}{Beam} &
\multicolumn{3}{c}{Separation} &
\colhead{$\delta v$}& 
 \\
\colhead{} &
\colhead{Size} &
\colhead{PA} &  
\colhead{$s_{\rm{CO-mm}}$} & 
\colhead{$s_{\rm{mm-opt}}$} & 
\colhead{$d_{\rm{mm-opt}}$} &
\colhead{}\\
\colhead{} &
\colhead{(\arcsec$\times$\arcsec)} & 
\colhead{($\deg$)} & 
\colhead{(\arcsec)} & 
\colhead{(\arcsec)}&
\colhead{(kpc)}&
\colhead{(km\,s$^{-1}$)}\\
\colhead{(1)} & 
\colhead{(2)} & 
\colhead{(3)} & 
\colhead{(4)} &
\colhead{(5)} &
\colhead{(6)} &
\colhead{(7)}
}
\startdata
DW001 & 3.9$\times$2.9 & -26 & $0.2\pm0.6$ NE & 0.4$\pm$0.5 W & 3$\pm$4 & 110$\pm$80 \\
DW002 & 3.5$\times$2.5 & 149 & $0.2\pm0.6$ SE & 1.6$\pm$0.5 S & 13$\pm$4 & 80$\pm$130 \\
DW003a & 6.2$\times$3.7 & 103 & - & 4.3$\pm$0.5 SE & 36$\pm$4 & 1480$\pm$120\\
DW003b &  &  & - & 10.6$\pm$1.0 NW & 88$\pm$8 & 1610$\pm$90 \\
DW004 & 3.5$\times$2.7 & -173 & - & - & - & - \\
DW005a & 3.5$\times$2.7 & 8 & $1.0\pm0.6$ S & 1.4$\pm$0.5 SW & 12$\pm$4 & 1420$\pm$130 \\
DW005b &  &  & - & 4.3$\pm$0.5 SE & 35$\pm$4 & 1160$\pm$90  \\
DW006 & 4.1$\times$2.6 & -33 & $0.6\pm0.6$ W & 0.4$\pm$0.5 W & 3$\pm$4 & 1810$\pm$80  \\
\enddata
\tablecomments{
(1) Sources names;
(2) the size and (3) positional angle (PA) of the clean beam; 
(4) the angular distance between the CO emission peak and the 2mm continuum peak;  
(5) the angular distance between the 2mm continuum peak and the optical quasar position; 
(6) the physical projected distance between the 2mm continuum peak and the optical quasar position; 
(7) the velocity difference between the CO emission line and optical redshifts with both optical and CO redshift uncertainties considered
(positive values indicate a redshifted CO emission compared to the optical lines).
Since DW003a and DW005b do not have SNR$>$5 continuum detection, 
we only list the separation between the CO peak and the optical positions in column (5) and (6). 
For DW003b, the spatial uncertainty is a combination of NOEMA pointing uncertainty and positional uncertainty.
For others, a $0.5\arcsec$ spatial uncertainty is the pixel scale.
For DW005a, the velocity offset is the average between the two peak positions, weighted by their relative fitted fluxes.}
\end{deluxetable*}

Column (2) in Table \ref{tab:off} lists the offsets of the peaks between the CO and dust continuum emissions ($s_{\rm{CO-mm}}$) with an error calculated from CO and continuum spatial detection uncertainties, NOEMA pointing accuracy and the pixel size.
For DW001 and DW002, 
their offsets are consistent 
within the astrometric accuracy of NOEMA ($0.2\arcsec$).
DW005a has a $1.0\arcsec$ positional offset,
but of different direction compared to DW005b (south vs. south-east),
thus likely not due to the latter.
Misaligned gas and dust components
suggest that DW005a itself may be in a pair system, 
while DW005b is a third component outside the pair.
Difference in the spatial distribution of the molecular gas and dust continuum 
has been observed with small separations at higher redshifts \citep[$\leq 1\arcsec$, e.g.][]{2021arXiv210903450G, 2022A&A...660A..60F, 2022arXiv220903380L}, 
though often times larger offsets are observed between the optical and submm components for both AGN/quasars and star forming galaxies (see Section \ref{sec:disoff}).

In general, we find spatial offsets of $\sim 1-4\,\arcsec$ 
between the $Herschel$ and the optical quasar positions.
We note that the positional errors are at $6\,\arcsec$ level, which propagate from the pointing accuracy \citep[$2\,\arcsec$, ][]{2010A&A...518L...1P}, 
spatial detection uncertainty ($\sim\,2\arcsec$), 
and the pixel size \citep[$6\arcsec$ at $250\mu$m,][]{2014ApJS..210...22V}. 
Given the large positional errors, 
the optical-FIR offsets are not discussed later.

At $z\sim2-3$, 1$\arcsec$ corresponds to $\sim$ 8\,kpc.
Thus for our targets,
the observed projected distance correspond to 3 to 13 kpc.
This offset is significant compared to the typical galaxy sizes at the Cosmic Noon \citep[a few kpc,][]{2020ARA&A..58..661F}.
For targets with a companion 2mm continuum component (DW003 and DW005),
their offsets to the quasar positions range from 35 to 90 kpc, 
suggesting that the second mm source is likely another galaxy.

Velocity shifts are also observed 
between the optical and mm spectroscopic redshifts (Table \ref{tab:off}).
The red dashed lines in Figure~\ref{fig:obs} (right) 
mark the expected frequencies and ranges based on the SDSS spectroscopic redshifts and associated uncertainties.
The differences between the optical and CO redshifts (i.e. $\delta v$)
are all positive,  corresponding to redshifted CO lines relative to the optical lines.
In DW001 and DW002, the velocity difference is relatively small 
with a large uncertainty and can be treated as consistent,
despite the relatively large spatial offset in DW002 (Column (5) and (6) in Table \ref{tab:off}).
The velocity difference is large (1100-1800 km\,s$^{-1}$) for DW003, DW005, and DW006.
We find significant velocity shifts, 
as compared to the redshift uncertainties,
in DW003a and DW005a.
These two systems may undergo volatile kinematic activities, 
possibly related to the secondary components. 
The origin and nature of these offsets will be further discussed in Section \ref{sec:disoff}.

\subsection{Molecular gas reservoir}
\label{sec:mgas}

We use the equation from \citet{1997ApJ...478..144S} to calculate CO(1-0) luminosity from the measured CO flux.
To convert the observed mid-J CO luminosity to CO(1-0) luminosity, $L'_{\rm{CO(1-0)}}$,
we adopt the conversion factors for quasars from \citet{2013ARA&A..51..105C},
i.e., 
r$_{41}$ = $L'_{\rm{CO(4-3)}}$/$L'_{\rm{CO(1-0)}}$ = 0.87,
and r$_{51}$ = $L'_{\rm{CO(5-4)}}$/$L'_{\rm{CO(1-0)}}$ = 0.69.

\begin{deluxetable*}{cccccc}[h]
\tablecaption{Molecular gas properties \label{tab:gas}}
\tablecolumns{5}
\tablewidth{0pt}
\tablehead{
\colhead{Source name} &
\colhead{$L'_{\rm{CO(4-3)}}$} & 
\colhead{$L'_{\rm{CO(5-4)}}$} & 
\colhead{$L'_{\rm{CO(1-0)}}$} &
\colhead{$M_{\rm{H_2}}$} \\
\colhead{}  &
\colhead{($10^{10} \rm{K~km~s^{-1}~pc^2}$)} 
& \colhead{($10^{10} \rm{K~km~s^{-1}~pc^2}$)} 
& \colhead{($10^{10} \rm{K~km~s^{-1}~pc^2}$)} 
& \colhead{($10^{10} M_{\rm{\odot}}$)}
}
\startdata
DW001  & - & 3.5$\pm$1.0 & 5.1$\pm$1.5 & 4.1$\pm$1.2 \\
DW002  & - & 8.0$\pm$1.9 & 11.6$\pm$2.7 & 9.2$\pm$2.1 \\
DW003a  & 4.6$\pm$1.1 & - & 5.3$\pm$1.3 & 4.2$\pm$1.0 \\
DW003b & 0.8$\pm$0.3 & - & 1.0$\pm$0.4 & 0.8$\pm$0.3 \\
DW004  & $<$1.0 & - & $<$1.2 & $<$1.0 \\
DW005a  & 7.4$\pm$1.3 & - & 8.5$\pm$1.5 & 6.8$\pm$1.2 \\
DW005b & 1.8$\pm$0.6 & - & 2.1$\pm$0.7 & 1.7$\pm$0.6 \\
DW006  & - & 8.3$\pm$1.0 & 12.1$\pm$1.4 & 9.7$\pm$1.2 \\
\enddata
\tablecomments{The factor $\mu$ is the possible gravitational amplification factor.
$M_{\rm{H_2}}$ is estimated from $L'_{\rm{CO(1-0)}}$ assuming a conversion factor $\alpha = 0.8 \,M_{\rm{\odot}}~\rm{(K~km/s~pc^2)}^{-1}$. The errors on $M_{\rm{H_2}}$ do not include the uncertainty on $\alpha$ and the uncertainty on the conversion to the ground transition.}
\end{deluxetable*}

To estimate the molecular gas mass, 
we adopt a linear relation between $L_{\rm{CO}}'$ and H$_2$ masses (i.e., $M_{\rm{H_2}} = \alpha L_{\rm{CO(1-0)}}'$),
assuming $\alpha = 0.8 \,M_{\rm{\odot}}~\rm{(K~km/s~pc^2)}^{-1}$ \citep{2022MNRAS.517..962D}.
For all detected targets (Table \ref{tab:gas}), 
this yields $M_{\rm{H_2}}$ of the order of 10$^{10}\, M_{\odot}$.
Note that these values are not corrected for possible magnification.
The uncertainties listed in the table are based on the line measurement,
without considering systematic uncertainties on the conversion factors.
For instance, if we adopted 
an $\alpha = 4.0\, M_{\rm{\odot}}~\rm{(K~km/s~pc^2)}^{-1}$, 
the resulting $M_{\rm{H_2}}$ would have be 5 times larger.
In addition, we note that the conversion factors between different CO lines from \citet{2013ARA&A..51..105C} 
are average values, with a large scatter \citep[e.g., a scatter of 0.5\,dex 
have been reported in
CO(4-3) to CO(1-0), see e.g.][]{2018MNRAS.479.1154B, 2014MNRAS.445.2599B, 2018A&A...612A..29B}.

\subsection{SED fitting}
\label{sec:sed}

The SEDs for the six sources include fluxes and upper limits from:
SDSS \citep{2015ApJS..219...12A,2020ApJS..249....3A},
the UKIRT Infrared Deep Sky Survey \citep[UKIDSS,][]{2007MNRAS.379.1599L},
the Wide Infrared Survey Explorer \citep[WISE,][]{2010AJ....140.1868W},
and HerS \citep[][see Table \ref{tab:prop}]{2014ApJS..210...22V}.
The measured 2mm flux density greater than 5$\sigma$ is used, 
otherwise a $5\sigma$ upper limit (Section \ref{sec:continuum}) is adopted.
We fit the SEDs of our HyLIRG-quasars using the CIGALE code \citep{2022ApJ...927..192Y}.
We adopt
an AGN component and a cold dust component in our SED fitting,
with the CO redshifts as input.
Table \ref{tab:cigale} lists the configuration for the fits.

We note that the FIR fluxes of DW003a in the HerS catalog 
may suffer from blending of both DW003a and DW003b (See Section \ref{sec:continuum}).
However, the current $Herschel$ resolution \citep[$\sim18\arcsec$ at 250$\mu$m,][]{2014ApJS..210...22V} 
makes it difficult to assign the flux.
Here we assign $31\%$ of the $Herschel$ fluxes to DW003a,
which corresponds to DW003a and b's flux ratio in the 2mm.
At 500$\mu$m, the beam size of SPIRE is $\sim36\arcsec$ \citep{2014ApJS..210...22V}.
Considering possible contamination from nearby sources,
we manually increase the error budget for $500\mu$m to 50$\%$in the fit. 

The properties derived from SED fitting are listed in Table \ref{tab:sed}. 
For DW001, DW002 and DW006,
the $\mu L_{\rm{IR, SB}}$ we calculate is consistent with 
the values in \citet{2016ApJ...824...70D} (Table \ref{tab:prop}) within $\sim 1\sigma$.
Their $\mu L_{\rm IR, SB}$ still satisfy the HyLIRG definition. 
However, if a magnification factor of 5-10 is applied for DW006,
as derived from Figure \ref{fig:hylirg} in Section \ref{sec:hylirg},
its intrinsic $L_{\rm{IR, SB}}$ will drop to ULIRG level.
For DW003a, in contrast to the $10^{13} L_{\odot}$ cold-dust IR luminosity reported by \citet{2016ApJ...824...70D}, 
the fitted $\mu L_{\rm IR, SB}$ is also in ULIRG level. 
In addition, the dust mass is of $10^{8} M_{\odot}$, 
which is only $\sim 1\%$ of molecular gas mass (Table \ref{tab:gas}).
This suggests a dusty companion source
that contributes to the majority of the 2mm emission.
The $\mu L_{\rm{IR, SB}}$ of DW004 is only $0.6\times10^{13} L_{\rm{\odot}}$,
about 50$\%$ lower than in \citet{2016ApJ...824...70D}.
For all sources except DW003a, the fitted dust mass is of the order of $10^{9}M_{\odot}$, 
indicating the dust-rich nature of these HyLIRG-quasars.

We then estimate the $\mu$SFRs from $\mu L_{\rm{IR, SB}}$
using the \citet{1998ApJ...498..541K} conversion after correcting for the Chabrier IMF.  
For all sources except DW003a and the undetected DW004, 
the calculated $\mu$SFR is $\sim 1700-2500 M_{\rm{\odot}}\,\rm{yr^{-1}}$ (Table \ref{tab:sed}), 
confirming the starburst nature of these HyLIRG-quasars.
We note that these $\mu L_{\rm IR, SB}$ and $\mu$SFRs may be upper limits in some cases.
Some quasars \citep[e.g. $\sim30\%$,][]{2017A&A...604A..67D} 
are found to heat cold dust in the host galaxies and 
contribute more than 40\% of the total infrared luminosities
 \citep[e.g.,][]{2015A&A...579A..60S, 2023MNRAS.518.3667D}.

\begin{deluxetable*}{ccccc}[h]
\tablecaption{CIGALE parameters for HyLIRG-quasar candidates \label{tab:cigale}}
\tablecolumns{5}
\tablewidth{0pt}
\tablehead{
\colhead{Module} &
\colhead{Parameter} & 
\colhead{Symbol} & 
\colhead{Values} 
}
\startdata
Cold dust emission: \citet{2014ApJ...780..172D} & Mass fraction of PAH & $q_{\rm{PAH}}$ & 0.47, 1.77, 3.19, 5.95, 7.32 \\ 
& Minimum radiation field & $U_{\rm{min}}$ & 0.1, 5.0, 10, 15, 20, 30, 40, 50 \\ 
& Power-law slope & $\alpha$ & 1.0, 1.5, 2.0, 2.5, 3.0 \\
& Luminated fraction & $\gamma$ & 0.10, 0.25, 0.50, 0.75, 0.99 \\
\hline
AGN emission: \citet{2006MNRAS.366..767F} & AGN fraction & $f_{\rm{AGN}}$ & 0.1, 0.2, 0.3, 0.4, 0.5, 0.6, 0.7, 0.8 \\
& Optical depth & $\tau$ & 0.1, 3.0, 10.0 \\
& Beta & $\beta$ & -1.0, -0.5, 0.0 \\
& Gamma & $\gamma$ & 0.0, 2.0, 4.0, 6.0 \\
& Angle between equatorial axis and line of sight ($^{\circ}$) & $\psi$ & 60.10, 70.10, 80.10, 89.99 \\
& Optical slope power-law index & $\delta$ & -1, -0.5, 0.0, 0.5, 1 \\
& Emissivity index & & 1.6, 1.8, 2.0, 2.2 \\
& Temperature of the polar dust (K) & & 80, 100, 120 \\
\enddata
\end{deluxetable*}

\begin{figure}[H]
\centering  

\subfigure{}{
\label{dw0012}
\includegraphics[align=c, width=0.45\textwidth]{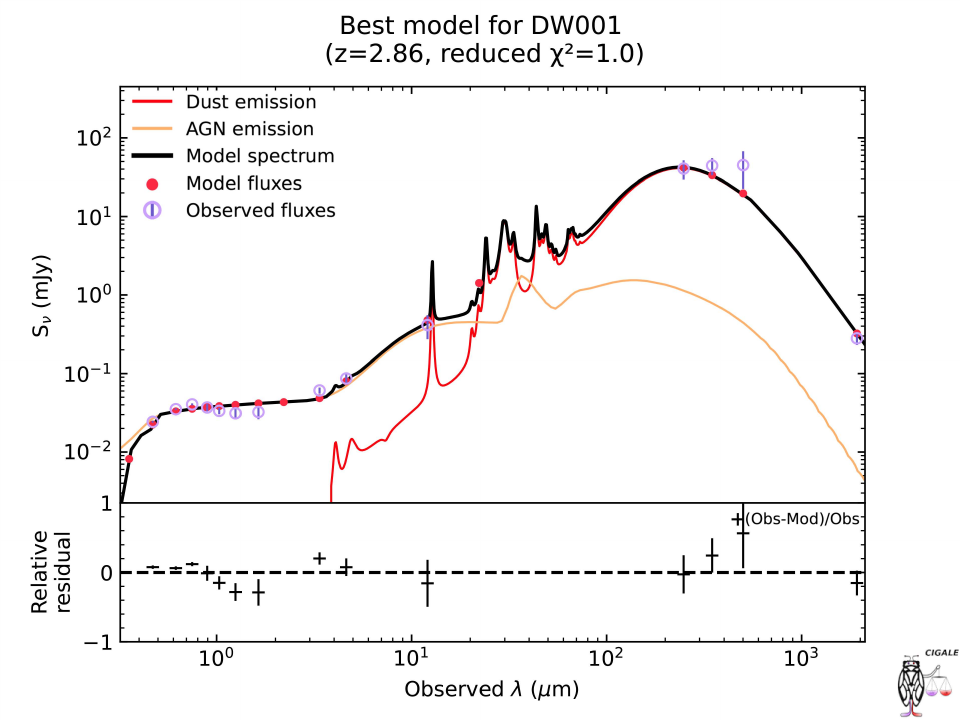}
\includegraphics[align=c, width=0.45\textwidth]{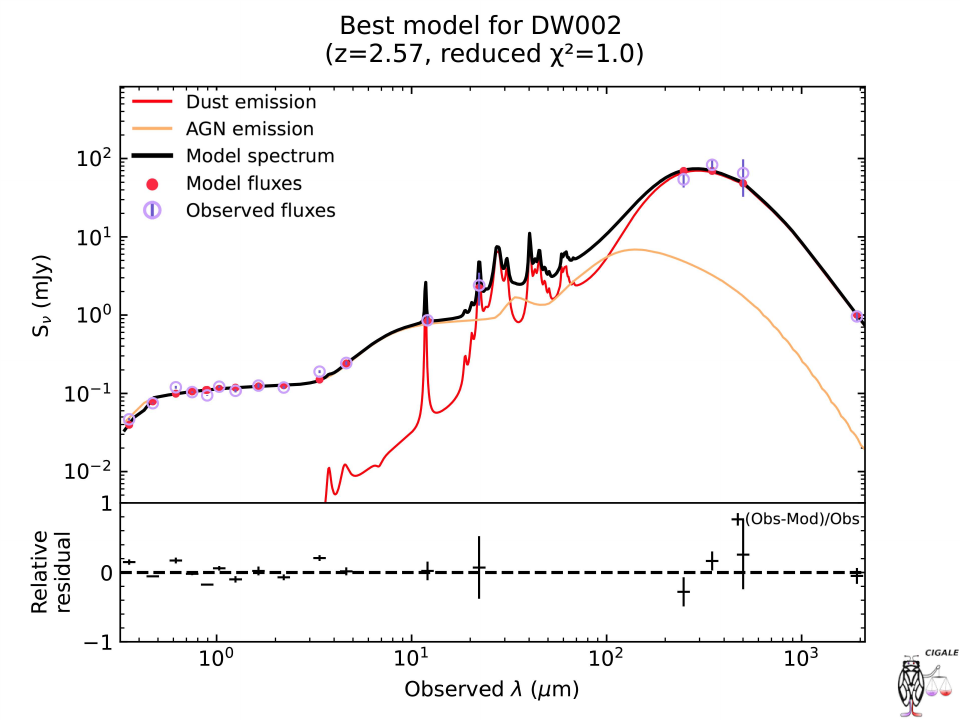}
}

\subfigure{}{ 
\label{dw0034}
\includegraphics[align=c, width=0.45\textwidth]{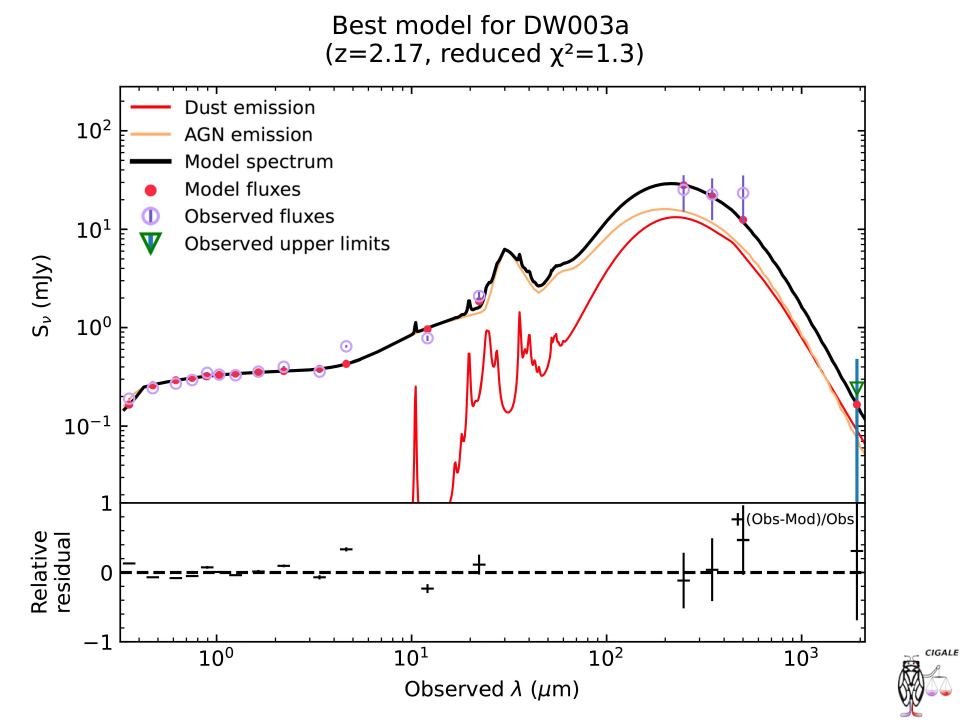}
\includegraphics[align=c, width=0.45\textwidth]{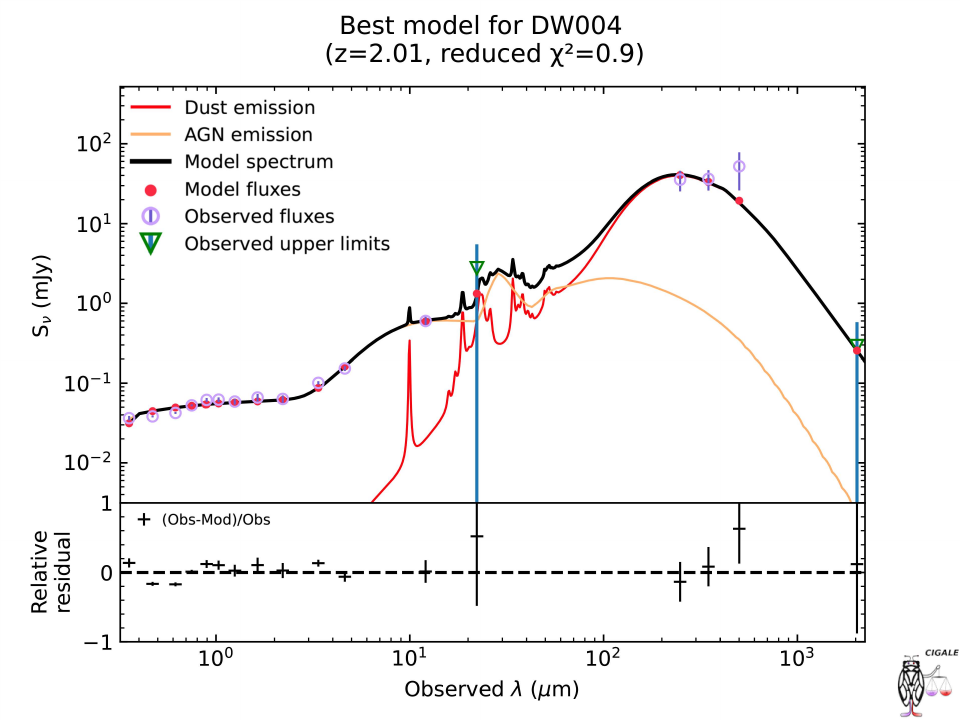}
}

\subfigure{}{
\label{dw0056}
\includegraphics[align=c, width=0.45\textwidth]{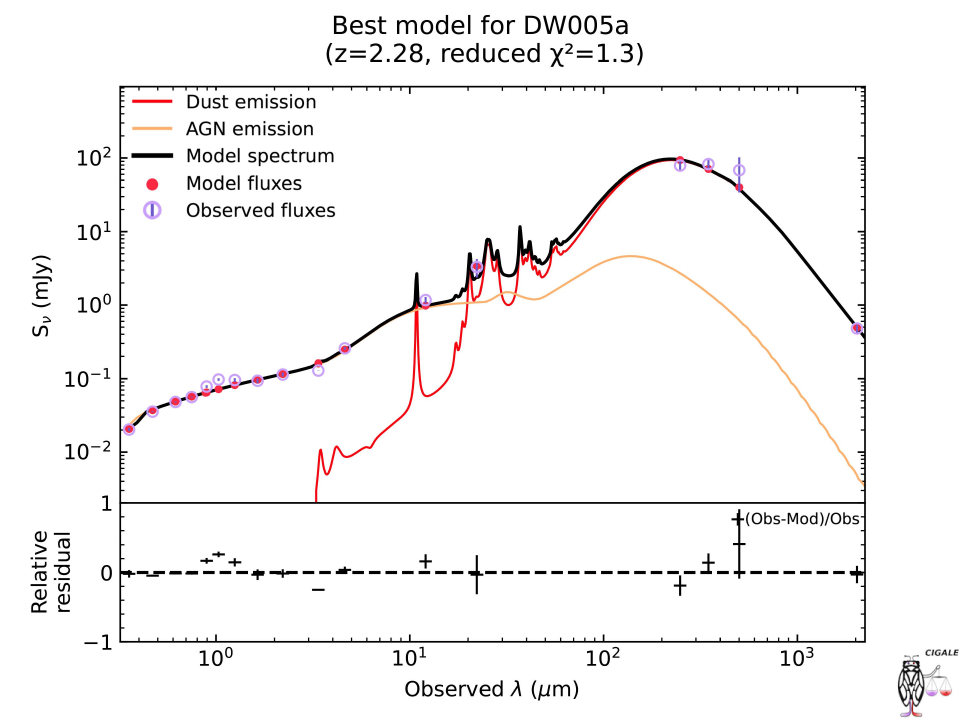}
\includegraphics[align=c, width=0.45\textwidth]{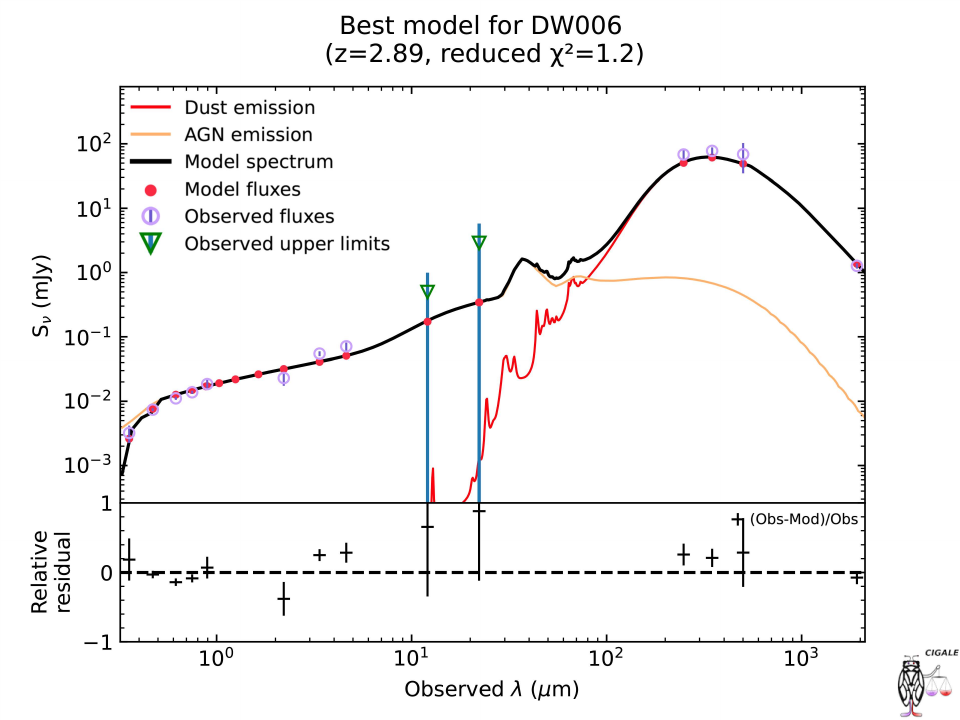}
}

\caption{SEDs of our HyLIRG-quasar sample. 
The FIR dust emission is  plotted in red, 
while the AGN component is in orange and 
in solid black is the composite SED. 
Violet empty circles mark the observed photometric data points, 
with upper limits shown in downward triangles.
Besides the NOEMA 2mm continuum fluxes, 
the data points are from SDSS, UKIDSS, WISE, $Herschel$, and ALMA (see Section \ref{sec:sed}).
}
\label{fig:sed}
\end{figure}

\newpage
\begin{deluxetable*}{cccccc}[ht]
\tablecaption{SED fitting results \label{tab:sed}}
\tablecolumns{4}
\tablewidth{0pt}
\tablehead{
\colhead{Name}  & 
\colhead{$f_{\rm IR, AGN}$} & 
\colhead{$\mu L_{\rm{IR, SB}}$} & 
\colhead{$\mu$SFR} &
\colhead{$\tau_{\rm{dep}}$} &
\colhead{$\mu M_{\rm{dust}}$} 
\\
\colhead{} & \colhead{} &
\colhead{($10^{13} L_{\odot}$)} &
\colhead{($M_{\rm{\odot}}\,\rm{yr^{-1}}$)} &
\colhead{(Myr)} &
\colhead{($10^8 M_{\odot}$)} 
}
\startdata
DW001 & 0.16$\pm$0.07 &1.7 $\pm$ 0.8 & 2000 $\pm$ 1000 & 20 $\pm$ 10 & 9 $\pm$ 2 \\
DW002 &  0.24$\pm$0.06 & 1.9 $\pm$ 0.6 & 2200 $\pm$ 700 & 40 $\pm$ 20 & 35 $\pm$ 8 \\
DW003a & 0.70$\pm$0.09 & 0.3 $\pm$ 0.1 & 400 $\pm$ 200 & 110 $\pm$ 50 & 7 $\pm$ 6 \\
DW004 & 0.25$\pm$0.07 & 0.6 $\pm$ 0.2 & 700 $\pm$ 300 & $<$10 & 15 $\pm$ 10 \\
DW005a & 0.11$\pm$0.03 & 2.1 $\pm$ 0.4 & 2500 $\pm$ 500 & 30 $\pm$ 10 & 21 $\pm$ 4 \\
DW006 & 0.10$\pm$0.01 & 1.4 $\pm$ 0.1 & 1700 $\pm$ 100 & 60 $\pm$ 10 & 48 $\pm$ 8
\enddata
\tablecomments{$f_{\rm IR, AGN}$ is fraction of AGN emission in total IR luminosity estimated from SED fitting with the CIGALE code.
$\mu L_{\rm{IR, SB}}$ is the AGN-removed, starburst-dominant IR luminosity. 
$\mu M_{\rm dust}$ is the fitted galactic dust mass.
The luminosities, SFRs and masses are apparent quantities not corrected for 
possible gravitational lensing effect.
$\tau_{\rm{dep}}$ is the depletion time calculated from Equation \ref{equ:dep}.}
\end{deluxetable*}

\section{Discussion} \label{sec:dis}
\subsection{HyLIRG diagnosis} \label{sec:hylirg}
\begin{figure}[H]
\centering  

\subfigure{}{
\includegraphics[width=1\textwidth]{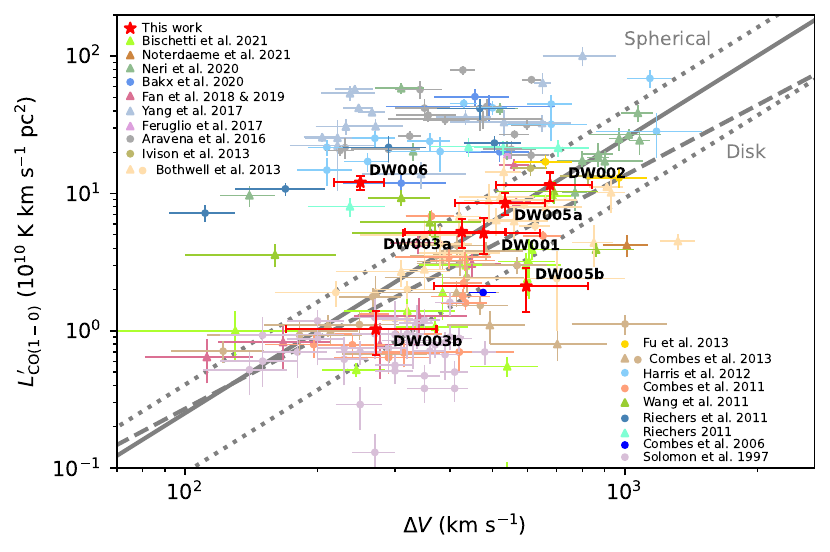}
}

\caption{ $L'_{\rm{CO(1-0)}}$ vs the CO FWHM ($\Delta\,V$) 
for our sample of 6 HyLIRG-quasars,
and 180 galaxies from the literature, 
including lensed and unlensed SMGs and DSFGs \citep{2013MNRAS.429.3047B,2016MNRAS.457.4406A,  2017A&A...608A.144Y,2012ApJ...752..152H,2020MNRAS.496.2372B}, 
local/low-to-mid-z ULIRGs \citep{1997ApJ...478..144S,2011A&A...528A.124C,2013A&A...550A..41C,2006A&A...460L..49C}, 
and high-z quasars and companions \citep{2018ApJ...856L...5F,2019ApJ...887...74F,2011ApJ...730..108R,2011ApJ...739L..32R,2010ApJ...714..699W,2017A&A...608A..30F,2021A&A...645A..33B,2021A&A...651A..17N}. 
Data points using CO high J transitions (J$\geq$3) are marked with triangles and converted 
to CO(1-0) using the same factors as in this paper \citep[i.e.,][]{2013ARA&A..51..105C}, 
while data points using CO J$\leq$2 transitions are marked with filled circles. 
The solid and dashed lines represents the approximate best-fitting quadratic relationships from \citet{2013MNRAS.429.3047B} and  \citet{2012ApJ...752..152H}, respectively.
The dotted lines represent the virial relations assuming spherical and disk models. 
}
\label{fig:hylirg}
\end{figure}

Similar to the optical images, no sign of gravitational lensing, e.g. arcs or filaments, is detected in the mm dust continuum down to the resolution of $1\arcsec$ of SDSS \citep{2016AJ....151...44D}.
Therefore, we refer to the lensing diagnosis diagram adapted from \citet{2012ApJ...752..152H},
based on the relation between the apparent CO luminosities, $L'_{\rm{CO(1-0)}}$, 
and the FWHM width, $\Delta$V, of the CO emission lines.
Figure \ref{fig:hylirg} presents our sample, along with the CO data of $180$ lensed and unlensed galaxies 
from the literature.
When applicable, 
we re-calculated the $L'_{\rm{CO(1-0)}}$ using the same cosmology and conversion factors 
adopted in this study for consistency (see Section \ref{sec:mgas}).

In Figure~\ref{fig:hylirg}, galaxies clearly fall in two distinct populations:
strongly lensed galaxies in the upper left and unlicensed along the power-law relation.
The unlensed or weakly lensed galaxies in general follow a virial relation:
\begin{equation}
L'_{\rm{CO(1-0)}} = \frac{C(\Delta V/2.355)^2R}{\alpha\cdot G}, 
\end{equation}
where $\Delta V$ is the FWHM of the CO line in km s$^{-1}$,
R is the radius of the CO emission region in parsecs, 
$\alpha$ is the conversion factor from $L'_{\rm{CO(1-0)}}$ to solar mass in $\rm{K\,km\,s^{-1}pc^2}$, 
G is the gravitational constant,
and C is a constant related to the kinematics of the galaxy.
We consider two extreme cases, using parameters suggested by \citet{2006ApJ...646..107E}: 
C = 2.1, R = 5 kpc, $\alpha$ = 4.6 for a disk model; 
C = 5, R = 2 kpc, $\alpha$ = 1.0 for a spherical model.
Both models are plotted with dotted lines in Figure \ref{fig:hylirg}, 
which nicely bracket the majority of the unlensed and slightly-lensed galaxies.

The solid line in Figure \ref{fig:hylirg} represents the 
best-fit relation $L'_{\rm{CO}} = 5.4\times \Delta V^2$ derived by \citet{2013MNRAS.429.3047B} and also applied by \citep[e.g.,][]{2015MNRAS.452.1140Z, 2020A&A...635A...7N}.
The dashed line represents the best-fit relation from \citet{2012ApJ...752..152H}, 
i.e., $L'_{\rm{CO(1-0)}} (\rm{K\,km\,s^{-1}pc^2}) = (\Delta V/400 \rm{km\,s^{-1}})^{1.7}\times10^{11}/3.5$. 
In our sample, four out of the five CO-detected sources 
are safely located in the unlensed or at most weakly lensed region.
For DW005a, whose CO emission shows double peaks, 
the separation between the two peaks was used as it's line width.
Despite the large error bars, 
the two possible companions, DW003b and DW005b, 
also fall in the unlensed region.

For objects with multiple CO lines observed \citep[e.g.][]{2013MNRAS.429.3047B},
we calculate the low-J (filled circle) and mid-J (triangle) emissions separately, 
and mark both in the diagram,
using $L'_{\rm{CO}}$/$L'_{\rm{CO(1-0)}}$ ratios from \citet{2013ARA&A..51..105C} for quasars and SMGs.
Detections with SNR$\leq$3 (tentative detections) are not plotted.

Based on the $L'_{\rm CO}$-$\Delta V$ diagram, 
DW001, DW002, DW003a, DW005a and two possible companions (DW003b, DW005b) 
fall in the unlensed power-law region, 
thus likely unlensed or at most weakly lensed, intrinsic HyLIRGs. 
DW004 is undetected thus not plotted.
DW006 is likely strongly lensed, with an estimated magnification factor of $\sim5-10$, 
based on the offset from the power-law relation,
making it a ULIRG system instead.  

\subsection{Depletion Time of the HyLIRGs}

Based on the gas mass and SFR calculated in Section~\ref{sec:mgas} and Section~\ref{sec:sed}, 
we estimate the gas depletion time:
\begin{equation}
\label{equ:dep}
{\tau_{\rm dep} = \frac {M_{\rm{H_2}}}{\rm{SFR}}}. 
\end{equation}
The calculated $\tau_{\rm{dep}}$(Table \ref{tab:sed}) ranges 
between $20-60$ Myr, 
similar to other starburst galaxies \citep[tens of Myrs, e.g.][]{2010ApJ...714L.118D, 2013A&A...550A..41C},
for all of our sources except DW003a.
This $\tau_{\rm{dep}}$ range is much shorter than 
the lifetime of a galaxy \citep[$\sim10$ Gyr for elliptical galaxies,][]{2006MNRAS.366..499D}. 
This is consistent with the scenario that HyLIRG-quasars are in a short transitional phase 
in the early formation of the massive elliptical galaxies \citep{2013Natur.498..338F}.
We also note that if a Salpeter IMF is applied, the SFR would also increase by $\sim0.15$ dex \citep[][]{2008MNRAS.385..147D}, 
resulting in even shorter $\tau_{\rm{dep}}$ of $\sim$15-40 Myr.
We also note that, as mentioned in Section \ref{sec:mgas}, 
the depletion time could be up to 5 times longer 
if instead of 0.8, 
a CO-H$_2$ conversion factor $\alpha$ of 4.0 $M_{\odot} ({\rm K\,km\,s^{-1}\,pc^2})^{-1}$ is adopted.

\begin{figure}[h]
\centering  

\subfigure{}{
\includegraphics[width=0.8\textwidth]{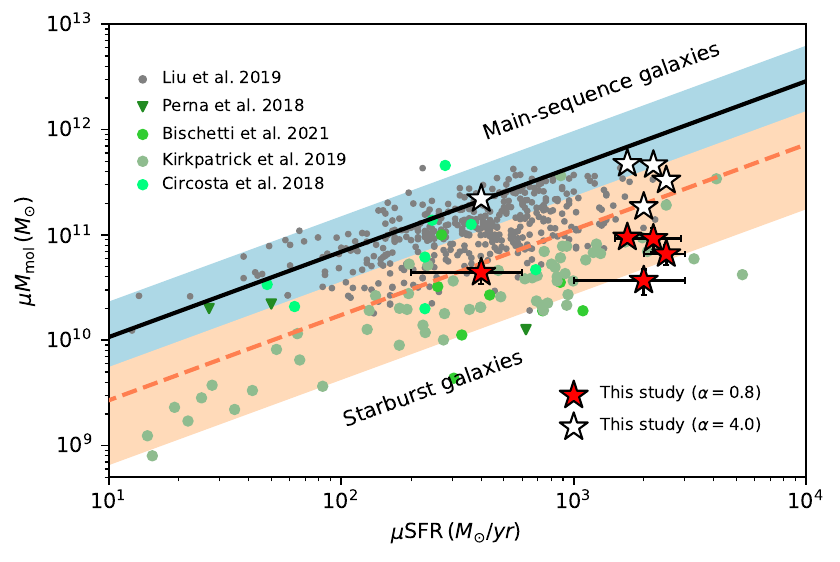}
}

\caption{The inverse integrated Kennicutt-Schmidt relation. 
The blue strip represents the relation for main-sequence (MS) galaxies, with a dispersion of 0.3 dex.
The solid black line is the MS relation consistent with Figure 3 in \citet{2020ARA&A..58..157T} at $z\sim2.5$.
The dashed orange line represents the MS relation offset by 0.6 dex to represents the starburst galaxies. 
The pink strip mark the starburst regions,
whose lower border represent the extreme starburst galaxies with a SFE $\sim15$ times higher 
than that of the MS galaxies \citep{2014ApJ...793...19S}.
Our five CO-detected quasars all fall in the starburst region, 
with a CO-H$_2$ conversion factor $\alpha = 0.8 M_{\odot} ({\rm K\,km\,s^{-1}\,pc^2})^{-1}$ (red stars),
similar to other starburst galaxies \citep{2020ARA&A..58..157T}.
They will fall in a transitional region between starburst and MS galaxies
if an $\alpha = 4.0 M_{\odot} ({\rm K\,km\,s^{-1}\,pc^2})^{-1}$ is adopted (white stars).
Grey dots are the star-forming galaxies 
in the COSMOS deep field from \citet{2019ApJ...887..235L} at z=2-3 plotted for comparison.
Filled circles and triangles (upper limits) are AGNs from 
the literature \citep{2018A&A...619A..90P, 2019ApJ...879...41K, 2021A&A...645A..33B, 2021A&A...646A..96C}.
No correction of gravitational magnification is applied to the sources.}
\label{fig:KS}
\end{figure}

In Figure~\ref{fig:KS},
we compare the inverse integrated Kennicutt-Schmidt relation between 
the molecular gas mass and $\mu$SFR of our sources, 
and observations from the literature.
Using the $\mu M_{\rm{H_2}}$ as a proxy for total molecular gas mass,
we find that the star formation efficiencies (SFEs) of all CO-detected sources follow the trend for starburst galaxies, similar to AGNs from literature.

\subsection{Origin of the Observed Positional and Velocity Offsets}\label{sec:disoff}

In Section \ref{sec:offset}, we present the spatial offset and velocity shifts between 
the optical quasars and the mm dust and molecular gas components.
In this section, we investigate the possible causes of these shifts.
Similar offsets have been ubiquitously observed in quasars and star forming galaxies, 
where the location of the molecular CO or dust components 
deviate from either the galaxies' optical positions, or redshifts, or both \citep[e.g.][]{2005A&A...439...75K, 2009ApJ...698L.188C, 2013A&A...550A..41C, 2016ApJ...829L..10I, 2017A&A...600A..57C, 2017ApJ...849L..36I, 2017A&A...603A..93M, 2018ApJ...866L...3B, 2021ApJ...923...59V}.

\begin{figure}[H]
\centering

\subfigure{}{
\label{civ0012}
\includegraphics[align=c, width=0.45\textwidth]{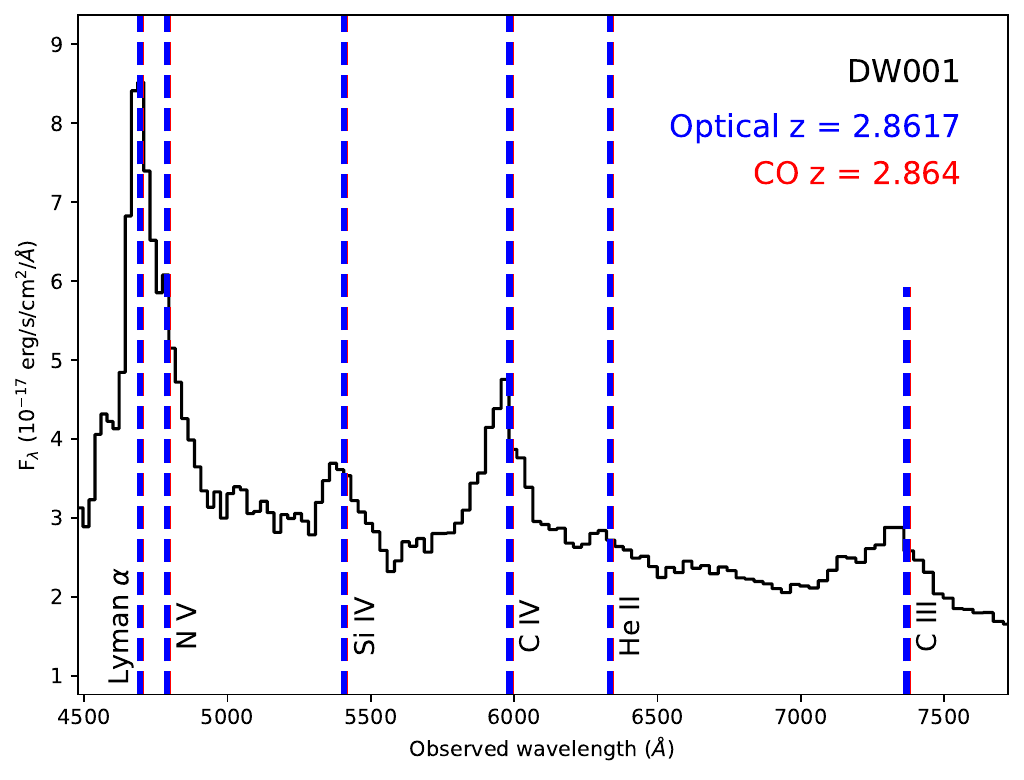}
\includegraphics[align=c, width=0.45\textwidth]{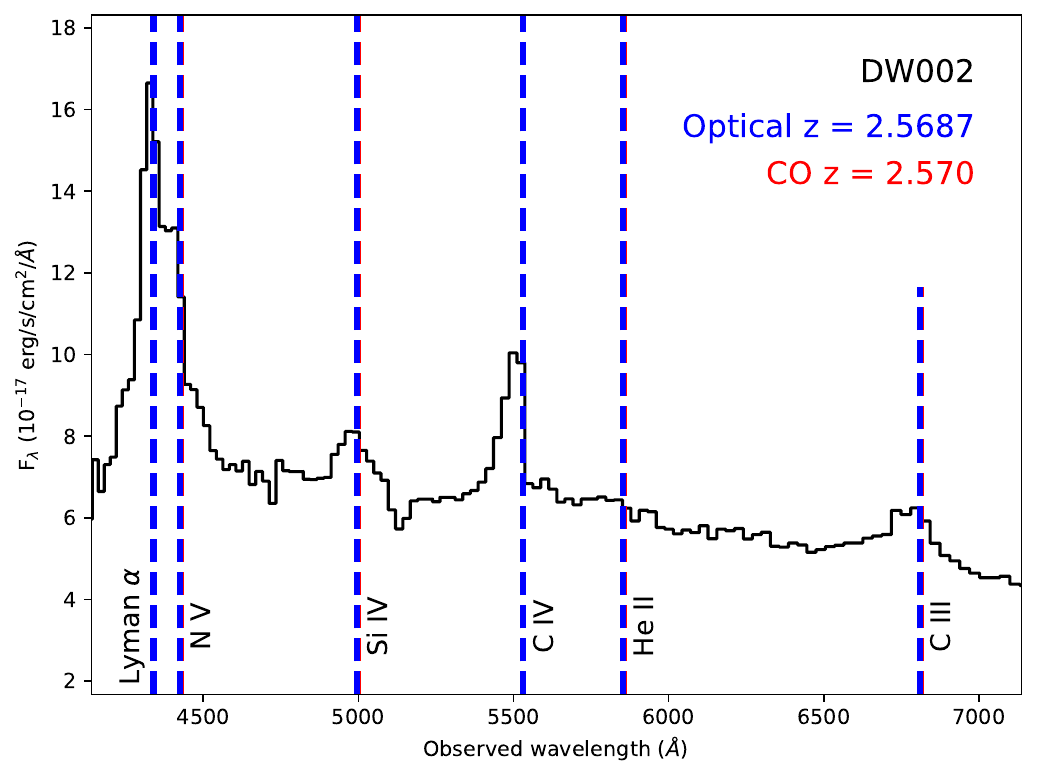}
}

\subfigure{}{ 
\label{civ0034}
\includegraphics[align=c, width=0.45\textwidth]{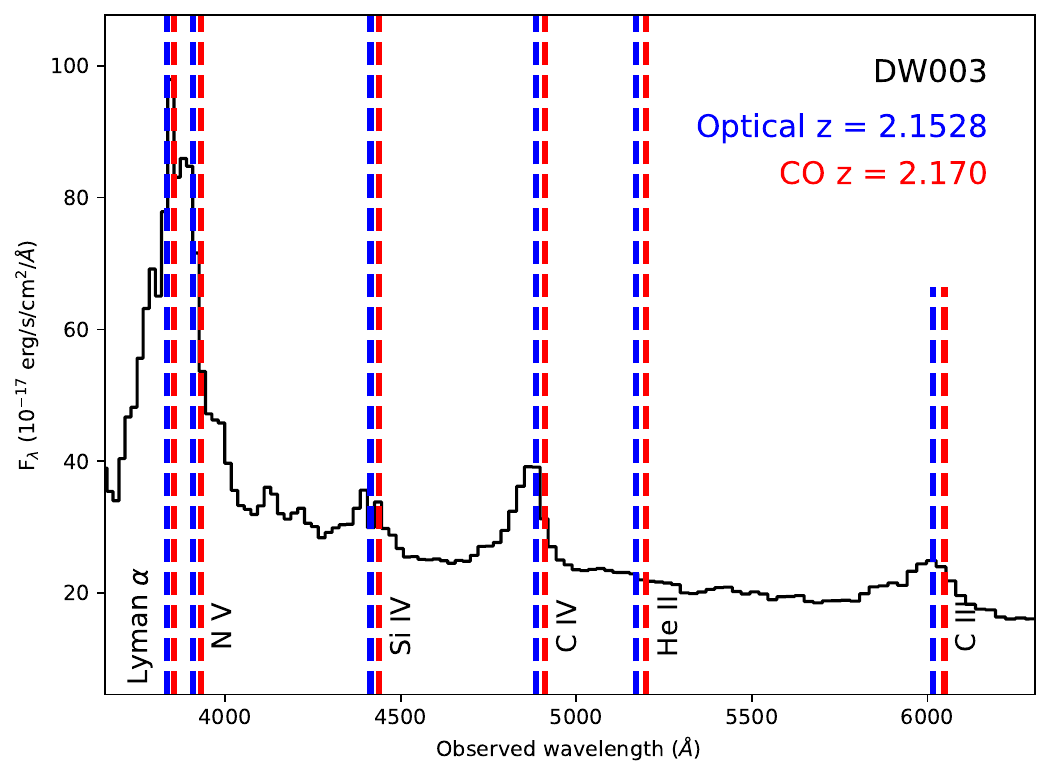}
\includegraphics[align=c, width=0.45\textwidth]{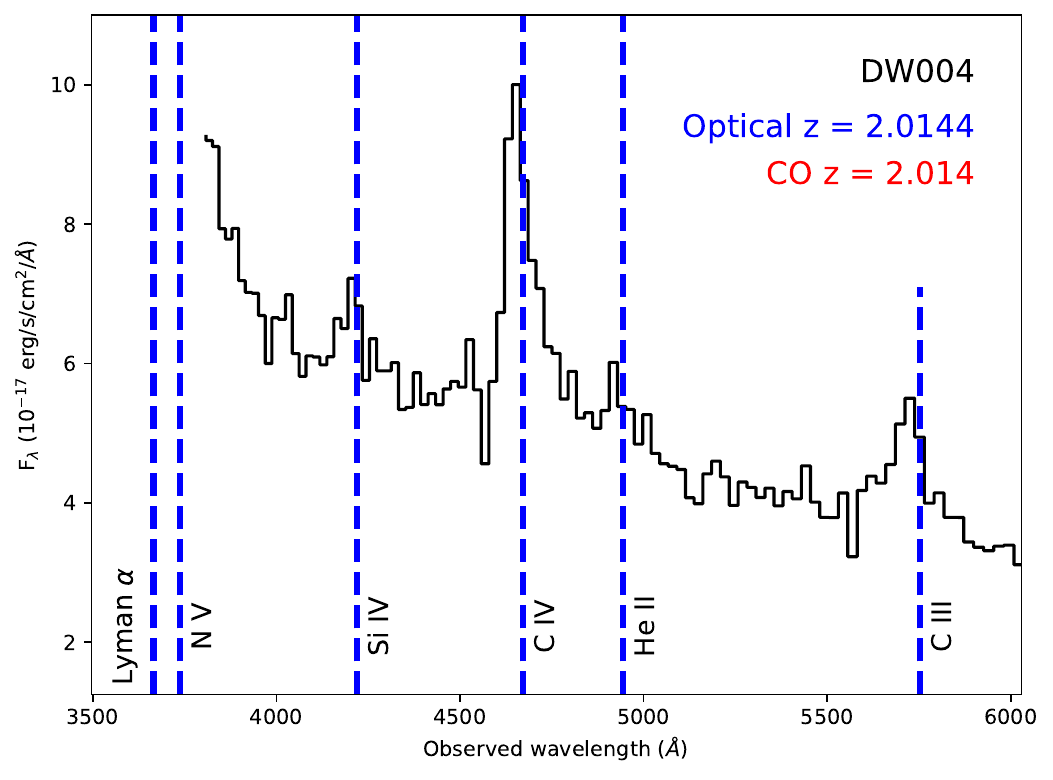}
}

\subfigure{}{ 
\label{civ0056}
\includegraphics[align=c, width=0.45\textwidth]{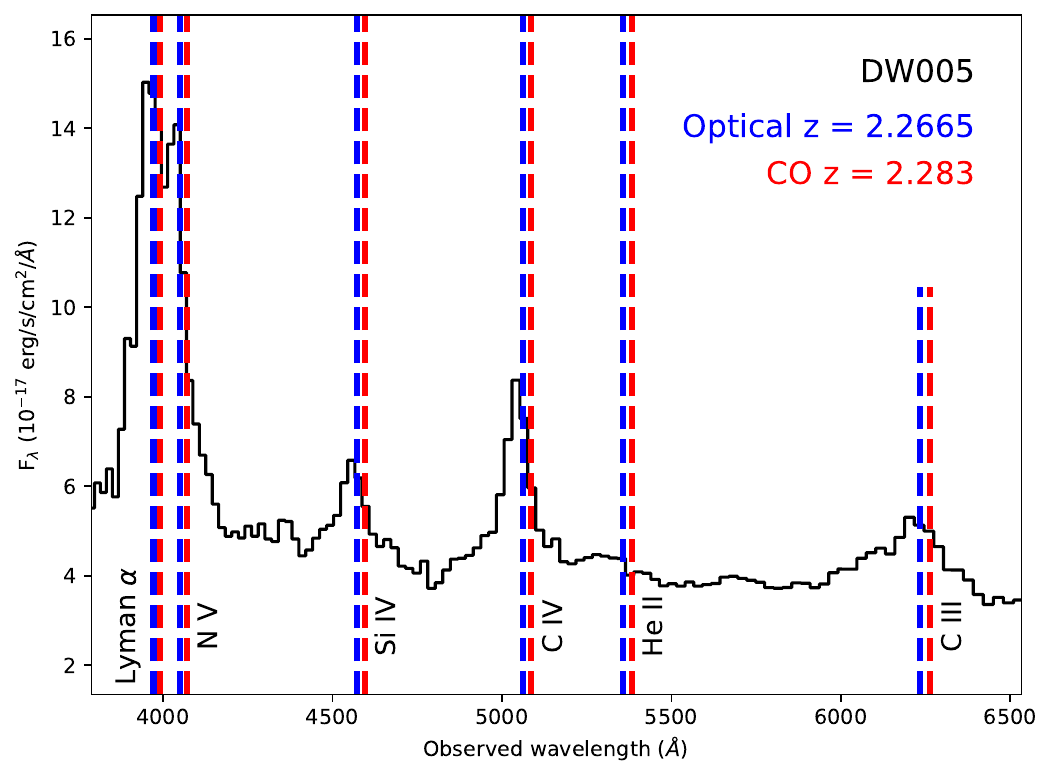}
\includegraphics[align=c, width=0.45\textwidth]{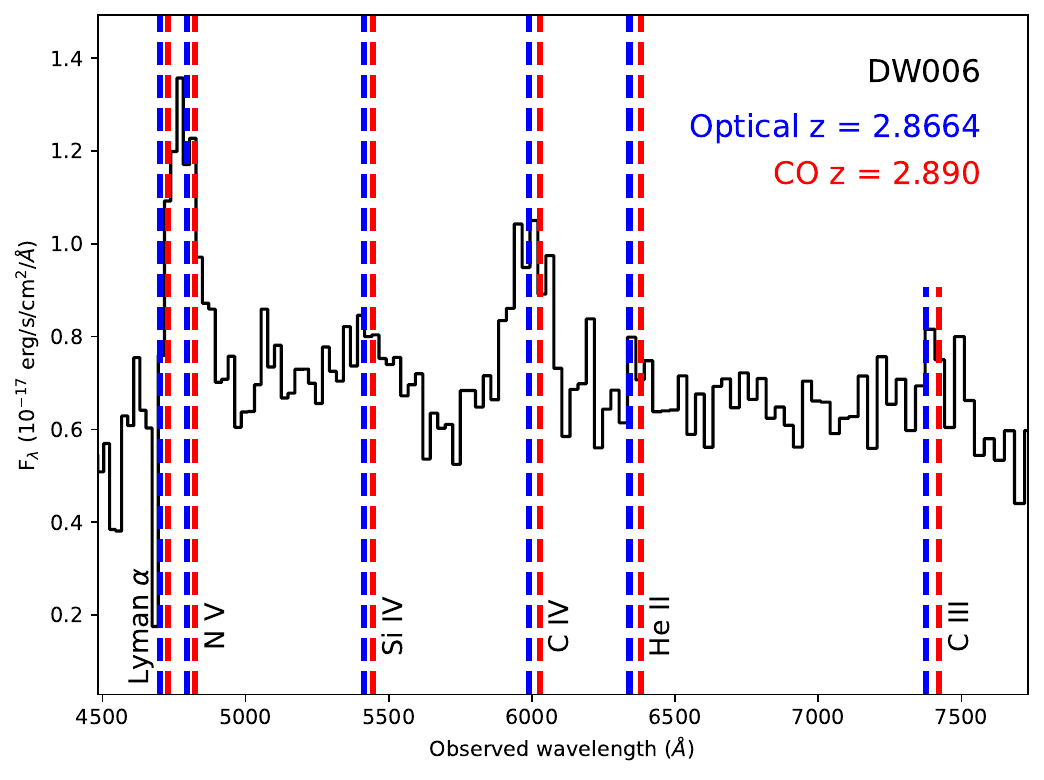}
}

\caption{Broad-emission-line spectra of DW001-DW006. The red dashed vertical lines mark the CO redshifts, and the blue dashed vertical lines mark the optical redshifts from SDSS (Table \ref{tab:prop}). For DW001 and DW002, the optical and CO redshifts are consistent and overlaid with each other. 
For DW003, DW005, DW006, the CO lines are redshifted from the peaks of optical lines.}
\label{fig:civ}
\end{figure}

One likely explanation for the systematic redshifted CO velocity 
is the known blueshift in the broad C IV emission lines in the quasar system,
whose peak is often blue-shifted with respect to narrow-line emissions, such as [O III], used as the representative systematic redshift of the quasar system when available \citep[see, e.g.,][]{2011AJ....141..167R, 2017MNRAS.465.2120C, 2020A&A...644A.175V}, and to cold gas tracers such as [C II] and CO \citep[see, e.g.,][]{2017A&A...598A.122B, 2017ApJ...836....8T, 2020ApJ...905...51S, 2021A&A...646A..96C}.
For our sources at z$\sim$2-3, indeed
the redshift determination from SDSS is highly reliable on the broad C IV broad emission lines,
partly due to the absence of [O III] or [Ne V] lines in the wavelength coverage. 
We double checked the spectra of DW002, DW003 and DW005,
whose Mg\,II broad emission lines are also well detected.
We find that all of the Mg\,II lines are $\lesssim 1000\,\rm{km\,s^{-1}}$ 
redshifted from the SDSS redshifts, 
fixed as the Gaussian center wavelength when fitting the Mg\,II line profiles.
If we treat the Mg\,II redshift as the quasar's representative redshift,
and remove any velocity difference from the CO-optical velocity offsets,
which now become -700, 600, and 500 $\rm{km\,s^{-1}}$ for DW002, DW003a and DW005, respectively.
 We note that the intrinsic blueshift of the broad-line-region could be 
as large as $\sim$2000 km s$^{-1}$  \citep[e.g.][]{2020ApJ...905...51S, 2018A&A...617A..81V},
which could explain the velocity shifts observed in our quasar sample.
A comparison of optical and CO redshifts on broad-line spectra is demonstrated in Figure \ref{fig:civ}.

Another scenario involves unresolved mergers in the HyLIRG-quasar systems, 
as in the cases in \citet{2005A&A...439...75K} and a number of similar targets \citep[i.e. a powerful AGN with a CO-rich merging partner, e.g.][]{2004ApJ...615L..17W,2005A&A...430L...1D}.
Merging galaxies with a dust-free quasar and an optically-obscured SMG
could explain the observed offsets, both positionally and spectroscopically, 
with the SMGs contributing the majority of the observed CO and dust emissions.
If two galaxies are in the later stage of a merger, 
they cannot be resolved under the current resolution ($\sim3.5\arcsec\times2.5\arcsec$),
even if the CO emission 
is centered at a second component slightly offset from the quasar location.
The SMG $+$ quasar scenario could also explain the 
observed velocity offsets up to several hundred of km s$^{-1}$.
The merging of two gas-rich galaxies can trigger starburst 
and explain the observed HyLIRG level luminosities,
as has been demonstrated in the cases of ULIRGs \citep[e.g.][]{2011ApJ...730..108R, 2019MNRAS.489..427I, huang23}.
In fact, DW003a has 
extended CO morphology, 
implying a second gas clump or galaxy 
at almost identical redshifts at a distance of $\sim$90\,kpc from the main component.
The double-peaked CO line profile of DW005a indicates disk-like rotation, or two distinct velocity components.
This offers indirect evidence to support the merger scenario.

Finally, we can not rule out the recoiling black holes scenario:
the central black hole being ejected with the broad line region (BLR) during galaxy merging, 
while the narrow lines still lag behind in the center of the galaxy, 
as well as the molecular gas \citep{2012AdAst2012E..14K}.
Velocity offsets generated by this effect can be of the order of 100-1000 km\,s$^{-1}$ \citep[as discussed in][]{2017A&A...600A..57C}, 
which is consistent with the observed velocity offset in our sample.
However, simulation suggests the probability of such an event is low \citep[only $<10\%$ of recoiling black holes formed from binary black holes have velocity $> 1000\, {\rm km\,s^{-1}}$,][]{2010ApJ...717..209C}.

\newpage
\section{Conclusion}

In this paper, we report NOEMA observations 
of a sample of six apparent HyLIRG-quasars,
selected as SDSS broad-emission-line quasars with Herschel IR emissions in the HyLIRG level.
Five out of the six quasars are detected or marginally detected 
in CO(5-4) or CO(4-3),
and four show well-detected dust continuum emissions.

The main conclusions of this paper are summarized as follows:
\begin{itemize}
\item{We confirm that out of the five CO detected quasars, 
four are consistent with being unlensed HyLIRGs,
based on their locations on the CO(1-0) luminosity ($L'_{\rm{CO(1-0)}}$) $vs$ linewidth ($\Delta V$) diagram.  
One source, DW006, is located well above the relation, 
suggesting a gravitationally amplified galaxy with $\mu\sim 5-10$.}

\item{Based on the cold molecular gas observations, 
we derive the molecular gas mass of our HyLIRG-quasars
to be $M_{\rm{H_2}} \sim$ $10^{10} M_{\rm{\odot}}$, 
with a depletion time scale of $\sim20-110$ Myr using a $M_{\rm{H_2}}/L'_{\rm{CO(1-0)}}$ conversion factor of 0.8 $M_{\rm{\odot}}\rm{(K~km/s~pc^2)}^{-1}$.
All of our CO-detected sources align well with starbursts.}

\item{The majority of our sources show significant positional (3-35\,kpc) offsets between the optical (quasar) and mm (dust/CO) emission peaks.
Most of them also have CO lines redshifted by $1000-2000\,\rm{km\,s^{-1}}$ in velocity 
compared to the optical redshifts,  likely related to blueshifted broad-line-regions observed in quasars.
The observed spatial offsets and velocity shifts are consistent with the scenario
that some of our HyLIRG-quasars are in a merger system with 
an optically obscured SMG and an optically-bright quasar.
The high SFR in these systems can be explained by the merger-triggered starburst.
}
\end{itemize}

Our HyLIRG-quasar sample offers a special population to investigate the connection between quasars and star formation activities in the most extreme starburst galaxies.
Their high molecular mass and likely unlensed nature indicate that these targets are in a fast transitional phase in between starburst and normal star-forming galaxies, while the common positional offset and spectroscopic velocity shifts indicate possible merging systems or complicated kinematics. 
High resolution observations, such as JWST and ALMA, will be helpful 
to identify more intrinsic HyLIRG-quasar systems, 
and to reveal their true nature.

The authors would like to thank Emmanuele Daddi and Shuowen Jin for 
constructive suggestions. 
This work is sponsored by the National Key R\& D Program of China (MOST) for grant No. \ 2022YFA1605300, 
the National Nature Science Foundation of China (NSFC) grants No.\ 12273051 and \ 11933003. 
Chentao Yang acknowledges support from an ERC Advanced Grant 789410.
Support for this work is also partly provided by the CASSACA.
This work is based on observations carried out under project number S20BT 
with the IRAM NOEMA Interferometer. 
IRAM is supported by INSU/CNRS (France), MPG (Germany) and IGN (Spain).

\end{document}